\documentclass[12pt]{article}
\usepackage[cp1251]{inputenc}
\usepackage{amsmath,amssymb,amsfonts,graphicx,indentfirst}
\usepackage[a4paper]{geometry}
\usepackage{caption2}
\usepackage {mathrsfs}
\usepackage {cite}

\begin{document}
\begin{center}{\large\bf Effects of anisotropy and Coulomb interactions on quantum transport in a quadruple quantum-dot structure}\end{center}
\begin{center}{M.\,Yu.\,Kagan$^{1,2}$, V.\,V.\, Val'kov$^{3}$, S.\,V.\, Aksenov$^{3}$}\end{center}
\begin{center}{kagan@kapitza.ras.ru, vvv@iph.krasn.ru, asv86@iph.krasn.ru}\end{center}
\begin{center}{\small$^1$ P.L. Kapitza Institute for Physical Problems RAS, 119334 Moscow, Russia}\end{center}
\begin{center}{\small$^2$ National Research University Higher School of Economics, 101000 Moscow, Russia}\end{center}
\begin{center}{\small$^3$ Kirensky Institute of Physics, Federal Research Center KSC SB RAS, 660036 Krasnoyarsk, Russia}\end{center}

\abstract{We present analytical and numerical investigation of spectral and transport properties of a quadruple quantum-dot (QQD) structure which is one of the popular low-dimensional systems in the context of fundamental quantum physics study, future electronic applications and quantum calculations. The density of states, occupation numbers and conductance of the structure were analyzed using the nonequilibrium Green's functions in the tight binding approach and the equation-of-motion method. In particular the anisotropy of hopping integrals and on-site electron energies as well as the effects of the finite intra- and interdot Coulomb interactions were investigated. It was found out that the anisotropy of the kinetic processes in the system leads to the Fano-Feshbach asymmetrical peak. We demonstrated that the conductance of QQD device has a wide insulating band with steep edges separating triple-peak structures if the intradot Coulomb interactions are taken into account. The interdot Coulomb correlations between the central QDs result in the broadening of this band and the occurrence of an additional band with low conductance due to the Fano antiresonances. It was shown that in this case the conductance of the anisotropic QQD device can be dramatically changed by tuning the anisotropy of on-site electron energies.}

\begin{center}\textbf{1. Introduction}\end{center}

Low-dimensional systems attract significant researchers' attention both with the possibility to study fundamental quantum physics and potential applications in nanoelectronics. One of the basic objects there are quantum dots (QDs). Different, often coexisting, processes such as the Kondo, Fano, Aharonov-Bohm effects as well as the Hubbard model physics are probed in the systems of QDs \cite{kikoin-01,oguri-05,delgado-08}. In the single-electron regime these structures are proposed to be utilized as spin qubits \cite{hanson-07,gimenez-09}. In addition, it was shown that they can act as rectifiers, spin filters, valves \cite{torio-04}.

Among QD-based structures, the arrays containing three and more QDs have been actively studying only recently due to more difficult experimental realization \cite{haider-09,hsieh-12}. The structures consisting of four QDs, QQD structures, were explored in different geometries. A nanosecond-timescale spin transfer of individual electrons across the serially connected QQD device was reported in \cite{baart-16}. In a square-like configuration such an operation was demonstrated on a closed path inside the QQD system \cite{thalineau-12}. In the same system with three electrons Nagaoka's ferromagnetism was observed \cite{stecher-10,barthelemy-13}, the features of the spin exchange of four electrons was studied \cite{mizel-04,scarola-05} and a self-contained quantum refrigerator was investigated \cite{venturelli-13}. It is important to emphasize that for all geometries the intra- and interdot Coulomb repulsion is a key factor influencing on the spectrum and transport properties \cite{nisikawa-06,ozfidan-13}.

\begin{figure}[h!] \begin{center}
\includegraphics[width=0.6\textwidth]{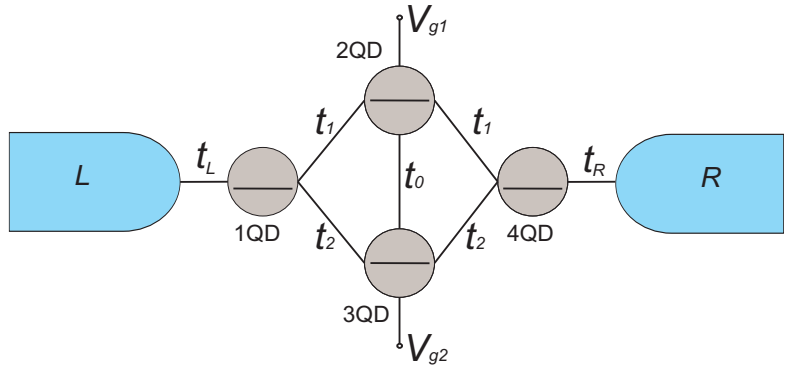}
  \caption{The QQD structure between metallic leads.} \label{model}
\end{center}
\end{figure}

The most common situation for quantum transport measurements of the QQD structure is when left and right metal contacts are coupled with two QDs so that other two QDs are situated in the central part (see fig. \ref{model}). The investigation of the Fano, Aharonov-Bohm and Aharonov-Casher interference effects in the Landauer formalism for this geometry earlier was restricted by the extreme cases of either strong Coulomb interaction (the Kondo regime) or the absence of it \cite{jiang-08,lobos-08}. Meanwhile, it was shown in \cite{lobos-08} that the QQD device subjected to the Rashba spin-orbit coupling acts as a spin filter. The similar behavior without the Rashba spin-orbit interaction and the Aharonov-Bohm effect was demonstrated for a multiple-QD network, the simplest case of which is the QQD \cite{fu-12}. In the last work the influence of the Coulomb interaction on the conductance was limited by the corresponding intradot term in the Hamiltonian. Thus the study of transport and spectral properties of the QQD structure in more general regime when both the finite intra- and interdot Coulomb interactions and the anisotropy effects are taken into account hasn't been considered yet. The anisotropy implies the difference of the hopping integrals in the QQD or on-site carrier energies (due to e.g. gate fields, $V_{g1}$ and $V_{g2}$) which takes place in experiments.

Here it is important to note that the introduction of the anisotropy allows us to effectively consider the QQD structure as the two-band Hubbard system. Let us remind that usually electron polaron effect (EPE) is present in multiband strongly correlated electron systems with substantially different electron bandwidths such as e.g. uranium based heavy fermion systems and other systems in mixed valence regime \cite{news-87,fulde-83,coleman-87,keitler-81} which can be described for example by the two-band Hubbard model with one narrow band (in case of sufficiently strong interband Hubbard interaction $V$) \cite{kagan-11a,kagan-11b} or Anderson Model (AM) \cite{anderson-70} with one-particle hybridization and two-particle Hubbard interaction between  s-p electrons of the light band and (heavy) electrons of localized f-d levels. In the two-band Hubbard model EPE is usually connected with the additional polaronic narrowing of the heavy particles bandwidth due to the dressing of the heavy particles in the virtual cloud of soft electron-hole pairs of the light particles. Similarly in the AM the EPE leads to the additional narrowing of the hybridization matrix element $t_{12}$. Note that in the QQD scheme (fig. \ref{model}) $t_{12}$ corresponds to the electron hopping $t_{2}$ from the  level in the shallow one-level trap in the left corner of the scheme (1QD) to the deep level in the  central trap (3QD) or correspondingly to the (reverse) hopping from the deep level in the central trap (3QD)to the shallow level in the trap (4QD), which is close to the right corner of the scheme at fig. \ref{model}. Consequently in all our calculations we should effectively replace $t_{2}$ by $t^{*}_{2} \ll t_{2}$ in case of strong EPE produced by large value of $V$. Thus in numerical analysis we will suppose that the anisotropy is induced by both specific design of the structure and the above-mentioned many body effects.

Note that the physics of EPE is closely connected  with the well-known results of Kondo, Nozieres et al on infrared divergences in the description of the Brownian motion of a heavy particle in a Fermi liquid of light particles \cite{kondo-64,iche-78} (see also important results of Yu. Kagan and N.V. Prokof’ev \cite{kagan-86,kagan-87}) as well as with the results on the infrared Mahan type \cite{mahan-67} divergences for the problem of X-ray photoemission from the deep electron levels \cite{nozieres-69} and with famous results of Anderson \cite{anderson-58} on the orthogonality catastrophe for the 1D chain of N electrons in the presence of one impurity in the system.

In all the cases both in uranium based heavy fermion systems \cite{kagan-11a,kagan-11b} and in other mixed valence systems such as manganite silicides, for example \cite{demishev-16}, EPE in the infinite geometry leads to anomalous resistivity characteristics in 3D and especially in 2D (layered) systems. Note that the manifestations of EPE are also very interesting in the restricted geometry of the microcontact (see the pioneering results of Matveev, Larkin \cite{matveev-92}) when we consider a charge sector of intradot Coulomb correlations \cite{arseev-92,arseev-12} and strong interdot Hubbard  correlations between electrons (or holes) of deep and shallow level in the central trap (2- and 3QDs). As we will show in this case, actual for the physics of dielectrics and molecularly doped polymers, we should have additional Fano-Feshbach \cite{fano-61,feshbach-62} many-body resonances in the tunneling conductance and an effective one-particle density of states for the microcontact. Note that the situation with small number of deep two-level traps randomly distributed between the large number of shallow one-level traps is described in the well-known paper of Bishop's group \cite{bishop-01} in connection with the physics of the radiation-induced conductivity (important for space applications) and more standard electron-phonon (or more exactly Holstein type configurational) polaron effects in molecularly doped polymers.

In this article, on the basis of the nonequilibrium Green's function technique and the tight binding approximation we studied spectral and transport characteristics of the QQD structure in which the intra- and interdot Coulomb correlations as well as the anisotropy effects take place. The interdot Coulomb interaction was considered between the QDs in the central part (see fig. \ref{model}). In order to define the role of the Coulomb interactions in the formation of the transport properties beyond the mean field approximation the decoupling scheme of You and Zheng \cite{you-99a,you-99b} was adapted.

The paper has been organized in six sections. The model Hamiltonian is described in Section 2. The nonequilibrium Green's function technique in the tight binding approximation is presented in Section 3. The analytical derivation of the retarded Green's functions of the QQD taking into account the Coulomb interactions is presented in Section 4. The results concerning the transport without the Coulomb interactions are presented in Section 5. The influence of the Coulomb interactions is considered in Section 6. Conclusions are given in Section 7.

\begin{center}\textbf{2. The model Hamiltonian}\end{center}

Let us consider electron quantum transport in the QQD structure depicted at figure \ref{model}. The system consists of three parts which are metallic leads and the structure between them. The Hamiltonian of the system
\begin{eqnarray} \label{Hfull}
\hat{H} = \hat{H}_L + \hat{H}_R + \hat{H}_{D} +\hat{H}_T.
\end{eqnarray}
The first two terms characterize the leads,
\begin{equation} \label{HLR}
\hat{H}_{L\left(R\right)}=\sum\limits_{k\sigma}\left(\xi_{k\sigma}\mp\frac{eV}{2}\right)
c^+_{L\left(R\right)k\sigma}c_{L\left(R\right)k\sigma},
\end{equation}
where $c^+_{L\left(R\right)k\sigma}$ is the creation operator in the left (right) lead with quantum number $k$, spin $\sigma$ and spin-dependent energy $\xi_{k\sigma}=\epsilon_{k\sigma}-\mu$; $\mu$ is the chemical potential of the system. It is supposed that the voltage $\pm V/2$ is applied to the left (right) lead.

The third term describes the QQD structure
\begin{eqnarray} \label{H4l}
&&\hat{H}_{D} =\sum\limits_{\sigma;j=1}^{4}\xi_{j\sigma}a^+_{j\sigma}a_{j\sigma}+U\sum\limits_{j=1}^{4}n_{j\uparrow}n_{j\downarrow}
+V\sum\limits_{\sigma\sigma'}n_{2\sigma}n_{3\sigma'}+\\
&&+\sum\limits_{\sigma}\left[t_{1}\left(a^+_{1\sigma}+a^+_{4\sigma}\right)a_{2\sigma}+
t_{2}\left(a^+_{1\sigma}+a^+_{4\sigma}\right)a_{3\sigma}+t_{0}a^{+}_{2\sigma}a_{3\sigma}+h.c.\right],~\nonumber
\end{eqnarray}
where $a_{j\sigma}$ annihilates the electron with spin $\sigma$ and energy $\xi_{j\sigma}=\epsilon_{j\sigma}-\mu$ on the $j$th QD; $t_{i}$, $i=0,~1,~2$ is a hopping matrix element between the QDs; $U$ is the intensity of the intradot Coulomb interaction; $V$ is the intensity of the interdot Coulomb interaction between the electrons in the central part (the $2$nd and $3$rd QDs).

The interaction between the leads and the QQD structure is determined by the last summand in \eqref{Hfull}
\begin{equation} \label{HT}
\hat{H}_T =t_{L}\sum \limits_{k\sigma}c_{Lk\sigma}^{+} a_{1\sigma} +
t_{R}\sum \limits_{k\sigma} c_{Rk\sigma}^{+} a_{4\sigma} + h.c.,
\end{equation}
where $t_{L\left(R\right)}$ is a hopping matrix element between the left (right) lead and the $1$st ($4$th) QD. Since the bias voltage is treated exactly it is convenient to perform a unitary transformation, $\hat{U}=\exp\left\{\frac{ieVt}{2}\sum\limits_{k\sigma}\left(n_{Rk\sigma}-n_{Lk\sigma}\right)\right\}$ \cite{rogovin-74}, to insert it in the tunnel Hamiltonian,
\begin{equation} \label{HT1}
\hat{H}_T =T_{L}\left(t\right)\sum\limits_{k\sigma}c_{Lk\sigma}^{+}a_{1\sigma} +
T_{R}\left(t\right)\sum\limits_{k\sigma}c_{Rk\sigma}^{+}a_{4\sigma} + h.c.,
\end{equation}
where $T_{L\left(R\right)}\left(t\right)=t_{L\left(R\right)}e^{\mp\frac{ieV}{2}t}$.

\begin{center}\textbf{3. Nonequilibrium Green's functions in the tight binding approach}\end{center}

To analyze transport properties of the system we utilize the nonequilibrium Green's functions method in the tight binding approximation \cite{keldysh-65,datta-95,datta-05}. Let us rewrite the Hamiltonian \eqref{Hfull} in terms of the operators $\widehat{\psi}_{L\left(R\right)k}$ and $\widehat{\psi}_{D}$,
\begin{equation} \label{psi}
\widehat{\psi}_{L\left(R\right)k}=\left(c_{L\left(R\right)k\uparrow}~c_{L\left(R\right)k\downarrow}\right)^T,~
\widehat{\psi}_{D}=\left(a_{1\uparrow}~a_{1\downarrow}~...~a_{4\uparrow}~a_{4\downarrow}\right)^T.\nonumber
\end{equation}
Then
\begin{equation} \label{HLRD1}
\hat{H}_{L\left(R\right)}=\sum\limits_{k}\widehat{\psi}_{L\left(R\right)k}^{+}
\widehat{\xi}_{k}\widehat{\psi}_{L\left(R\right)k},~
\hat{H}_{D} = \widehat{\psi}_{D}^{+}\widehat{h}_{D}\widehat{\psi}_{D},
\end{equation}
\begin{equation} \label{HT2}
\hat{H}_{T} =T_{L}\left(t\right)\sum\limits_{k}\widehat{\psi}_{Lk}^{+}\widehat{P}_{1}\widehat{\psi}_{D}+
T_{R}\left(t\right)\sum\limits_{k}\widehat{\psi}_{Rk}^{+}\widehat{P}_{4}\widehat{\psi}_{D}+h.c.,
\end{equation}
where
\begin{eqnarray}
&&\widehat{h}_{D}=
\left(\begin{array}{cccc}
\widehat{\xi}_{1} & \widehat{t}_{1} & \widehat{t}_{2} & \widehat{0} \\
\widehat{t}_{1} & \widehat{\xi}_{2} & \widehat{t}_{0} & \widehat{t}_{1} \\
\widehat{t}_{2} & \widehat{t}_{0} & \widehat{\xi}_{3} & \widehat{t}_{2} \\
\widehat{0} & \widehat{t}_{1} & \widehat{t}_{2} & \widehat{\xi}_{4}
\end{array}\right),~\label{hLRD1}\\
&&\widehat{t}_{i}=diag\left(t_{i}\right),~\widehat{\xi}_{l}=diag\left(\xi_{l\uparrow},~
\xi_{l\downarrow}\right),~l=k,1,...,4.\label{hLRD2}
\end{eqnarray}
The operators $\widehat{P}_{1}=\left(\widehat{I}~\widehat{0}\right)$ and $\widehat{P}_{4}=\left(\widehat{0}~\widehat{I}\right)$ project matrices on the subspace related to the $1$st or $4$th QD respectively. They consist of the $2 \times 2$ unitary matrix, $\widehat{I}$, and the zero block, $\widehat{0}$.

An electrical current operator in the left lead is determined by the corresponding charge change per time unit, $\widehat{I}_{L}=e\dot{N}_{L}$, where $N_{L}=\sum_{k\sigma}c_{Lk\sigma}^{+}c_{Lk\sigma}$ is the carrier number operator in the left lead. Using the equation of motion for Heisenberg operators $\left\langle\widehat{I}_{L}\right\rangle$  becomes
\begin{equation} \label{IL1}
\left\langle\widehat{I}_{L}\right\rangle \equiv I_{L}=ie\sum_{k}\Biggl\langle T_{L}^{+}\widehat{\psi}_{D}^{+}\widehat{P}_{1}^{+}\widehat{\psi}_{Lk}
-T_{L}\widehat{\psi}_{Lk}^{+}\widehat{P}_{1}\widehat{\psi}_{D}\Biggr\rangle.
\end{equation}
Let us introduce the nonequilibrium matrix Green's functions as
\begin{equation} \label{Gnm} \widehat{G}_{nm}^{ab}\left(\tau,~\tau'\right)=-i\left\langle\hat{T}_{C}\widehat{\psi}_{n}\left(\tau\right)\otimes
\widehat{\psi}_{m}^{+}\left(\tau'\right)\right\rangle,~n,~m=k,~D.
\end{equation}
Their time evolution is considered on the Keldysh contour, $C$. The indexes $a,~b=+,~-$ denote the branches of the Keldysh contour, $C_{+}$ and $C_{-}$. Then the current is expressed as
\begin{equation} \label{IL2}
I_{L}=2e\sum_{k}Tr\Biggl[Re\Bigl\{T_{L}^{+}\left(t\right)
\widehat{G}_{Lk,1}^{+-}\left(t,~t\right)\Bigr\}\Biggr],~
\end{equation}
where $\widehat{G}_{Lk,1}^{+-}\left(t,~t\right)=-i\Biggl\langle\hat{T}_{C}\widehat{\psi}_{Lk}\left(t\right)\otimes
\Bigl(\widehat{P}_{1}\widehat{\psi}_{D}\left(t\right)\Bigr)^{+}S_{C}\Biggr\rangle_{0}$ is a mixed lesser Green's function. In the last definition the averaging is made over the states of the system without interaction \eqref{HT2}. As a result the scattering matrix, $S_{C}=\hat{T}_{C}\exp\left\{-i\int_{C}d\tau\hat{H}_{T}\left(\tau\right)\right\}$, appears. Since the Hamiltonian of the device, $\hat{H_{D}}$, is formally the free-particle one, the rules for the second quantization operators can be utilized at the diagrammatic expansion of $\widehat{G}_{Lk,1}^{+-}\left(t,~t\right)$. Hence the current is written as
\begin{equation} \label{IL3}
I_{L}=2e\int_{C}d\tau_{1}Tr\Biggl[Re\Bigl\{\widehat{\Sigma}_{L}^{+a}\left(t-\tau_{1}\right)
\widehat{P}_{1}\widehat{G}_{D}^{a-}\left(\tau_{1}-t\right)\widehat{P}_{1}^{+}\Bigr\}\Biggr],~
\end{equation}
where $\widehat{\Sigma}_{L}^{ab}\left(\tau-\tau'\right)=T_{L}^{+}\left(\tau\right)
\widehat{g}_{Lk}^{ab}\left(\tau-\tau'\right)T_{L}\left(\tau'\right)$ is the self-energy function characterizing the influence of the left lead on the structure; $\widehat{g}_{Lk}^{ab}\left(\tau-\tau'\right)$ is the one-electron Green's function of the left lead. Taking into account the relations $\widehat{G}_{nm}^{--}=\widehat{G}_{nm}^{+-}-\widehat{G}_{nm}^{a}$, $\widehat{\Sigma}_{L}^{++}=\widehat{\Sigma}_{L}^{r}+\widehat{\Sigma}_{L}^{+-}$ (the indexes "r, a"$~$mean "retarded"$~$and "advanced"$~$correspondingly) and using the Fourier transform we obtain
\begin{equation} \label{IL4}
I_{L}=2e\int\limits_{-\infty}^{+\infty}\frac{d\omega}{2\pi}Tr\Biggl[Re\Bigl\{
\widehat{\Sigma}_{L}^{r}\widehat{P}_{1}\widehat{G}_{D}^{+-}\widehat{P}_{1}^{+}+
\widehat{\Sigma}_{L}^{+-}\widehat{P}_{1}\widehat{G}_{D}^{a}\widehat{P}_{1}^{+}\Bigr\}\Biggr].
\end{equation}
The Dyson and Keldysh equations for the full retarded and lesser Green's functions of the structure are
\begin{eqnarray}
&&\widehat{G}^{r}=\left(\left(\omega+i\delta\right)\widehat{I}-\widehat{h}_{D}-\widehat{P}_{1}^{+}\widehat{\Sigma}_{L}^{r}\widehat{P}_{1}-
\widehat{P}_{4}^{+}\widehat{\Sigma}_{R}^{r}\widehat{P}_{4}\right)^{-1},~\widehat{G}^{a}=\left(\widehat{G}^{r}\right)^{+}\label{Gr}\\
&&\widehat{G}^{+-}=\widehat{G}^{r}\left(\widehat{P}_{1}^{+}\widehat{\Sigma}_{L}^{+-}\widehat{P}_{1}+
\widehat{P}_{4}^{+}\widehat{\Sigma}_{R}^{+-}\widehat{P}_{4}\right)\widehat{G}^{a}.~\label{G+-}
\end{eqnarray}
The retarded and lesser self-energy functions are given by
\begin{equation} \label{Sigma}
\widehat{\Sigma}_{L\left(R\right)}^{r}=-\frac{i}{2}diag\left(\Gamma_{L\left(R\right)\uparrow},~\Gamma_{L\left(R\right)\downarrow}\right),~
\widehat{\Sigma}_{L\left(R\right)}^{+-}=i f\left(\omega \pm \frac{eV}{2}\right)diag\left(\Gamma_{L\left(R\right)\uparrow},~\Gamma_{L\left(R\right)\downarrow}\right),
\end{equation}
where $\Gamma_{L\left(R\right)\sigma}\left(\omega\right)=\pi t_{L\left(R\right)}^2\rho_{L\left(R\right)\sigma}\left(\omega\right)$ is the coupling strength between the structure and the left (right) lead characterized by its density of states $\rho_{L\left(R\right)\sigma}\left(\omega\right)=\sum\limits_{k}\delta\left(\omega-\xi_{k\sigma}\right)$; $f\left(\omega \pm \frac{eV}{2}\right)$ is the Fermi distribution function. In this study the leads are paramagnetic metals and treated in the wide-band limit, i.e. $\Gamma_{L\left(R\right)\sigma}=const$. After substitution \eqref{Sigma} into \eqref{IL4} and using the relation $i\left(\widehat{G}_{D}^{r}-\widehat{G}_{D}^{a}\right)=\widehat{G}_{D}^{r}\left(\Gamma_{L}\widehat{P}_{1}^{+}\widehat{P}_{1}+
\Gamma_{R}\widehat{P}_{4}^{+}\widehat{P}_{4}\right)\widehat{G}_{D}^{a}$ \cite{datta-05} the final general expression describing the current can be written as follows
\begin{equation} \label{IL}
I_{L}=e\int\limits_{-\infty}^{+\infty}\frac{d\omega}{2\pi}Tr\Bigl[\widehat{T}\left(\omega\right)\Bigr]
\Bigl(f\left(\omega - eV/2\right)-f\left(\omega + eV/2\right)\Bigr),
\end{equation}
where $\widehat{T}\left(\omega\right)=\Gamma_{L}\Gamma_{R}\widehat{G}_{14}^{r}\left(G_{14}^{r}\right)^{+}$ is the matrix transmission coefficient; $\widehat{G}_{14}^{r}=\widehat{P}_{1}\widehat{G}_{D}^{r}\widehat{P}_{4}^{+}$.

In further numerical calculations the system will be considered at low temperatures. Moreover, in this study we will be interested in the behavior of the differential conductance, $G=dI_{L}/dV$, as a function of the gate voltage, $\epsilon_{D}$ (hereinafter we suppose that $\epsilon_{j\sigma}=\epsilon_{D}$), at low bias (so called linear regime). Consequently, expanding $f\left(\omega \pm eV/2\right)$ into the Taylor series and taking into account $-df\left(\omega\right)/d\omega \approx \delta\left(\omega\right)$ we get the Landauer-Buttiker formula
\begin{equation} \label{G}
G=G_{0}Tr\Bigl[\widehat{T}\left(\epsilon_{D},~\omega=0\right)\Bigr],
\end{equation}
where $G_{0}=e^2/h$ is the conductance quantum. The total density of states (TDOS) is given by
\begin{equation}\label{DOS}
\rho=\frac{i}{2\pi}Tr\left[G_{D}^{r}-G_{D}^{a}\right].
\end{equation}

\begin{center}\textbf{4. The retarded Green's function of the QQD structure with the Coulomb interactions}\end{center}

In this section we describe the effects of Coulomb interactions on the transport properties of the QQD structure. In order to achieve this we employ the equation-of-motion technique for the retarded Green's functions, $G_{i\sigma j\sigma'}^{r}\left(\omega\right)=\langle\langle a_{i\sigma} | a_{j\sigma'}^{+} \rangle\rangle$, which are the Fourier transform of $G_{i\sigma j\sigma'}^{r}\left(t,~t'\right)=-i\Theta\left(t-t'\right)\left\langle \left\{a_{i\sigma}\left(t\right),~a_{j\sigma'}^{+}\left(t'\right)\right\}\right\rangle$. The equation for $G_{i\sigma j\sigma'}^{r}\left(\omega\right)$ is
\begin{equation} \label{eqG1}
z\langle\langle a_{i\sigma} | a_{j\sigma'}^{+} \rangle\rangle=\left\langle\left\{a_{i\sigma},~a_{j\sigma'}^{+}\right\}\right\rangle+
\langle\langle \left[a_{i\sigma},~\hat{H}\right] | a_{j\sigma'}^{+} \rangle\rangle,
\end{equation}
where $z=\omega+i\delta$ and $\hat{H}$ has the form \eqref{Hfull}. Since the 2nd and 3rd QDs are identical in the considered system we denote them by the indexes $\alpha$ and $\overline{\alpha}$. The indexes of the 1st and 4th QDs are $\beta$ and $\overline{\beta}$ for the same reason. As a result the equation for $\langle\langle a_{\alpha\sigma} | a_{\alpha\sigma}^{+} \rangle\rangle$, $\langle\langle a_{\beta\sigma} | a_{\alpha\sigma}^{+} \rangle\rangle$ and $\langle\langle c_{L\left(R\right)k\sigma} | a_{\alpha\sigma}^{+} \rangle\rangle$ are
\begin{eqnarray}
&&\left(z-\xi_{\alpha}\right)\langle\langle a_{\alpha\sigma} | a_{\alpha\sigma}^{+} \rangle\rangle=1+U\langle\langle n_{\alpha\overline{\sigma}}a_{\alpha\sigma} | a_{\alpha\sigma}^{+} \rangle\rangle+\nonumber\\
&&+V\left(\langle\langle n_{\overline{\alpha}\sigma}a_{\alpha\sigma} | a_{\alpha\sigma}^{+} \rangle\rangle+\langle\langle n_{\overline{\alpha}\overline{\sigma}}a_{\alpha\sigma} | a_{\alpha\sigma}^{+} \rangle\rangle\right)+t_{0}\langle\langle a_{\overline{\alpha}\sigma} | a_{\alpha\sigma}^{+} \rangle\rangle+\nonumber\\
&&+t\left(\alpha\right)\left(\langle\langle a_{\beta\sigma} | a_{\alpha\sigma}^{+} \rangle\rangle+\langle\langle a_{\overline{\beta}\sigma} | a_{\alpha\sigma}^{+} \rangle\rangle \right),\label{eqG2}\\
&&\left(z-\xi_{\beta}\right)\langle\langle a_{\beta\sigma} | a_{\alpha\sigma}^{+} \rangle\rangle=U\langle\langle n_{\beta\overline{\sigma}}a_{\beta\sigma} | a_{\alpha\sigma}^{+} \rangle\rangle+t\left(\alpha\right)\langle\langle a_{\alpha\sigma} | a_{\alpha\sigma}^{+} \rangle\rangle+\nonumber\\
&&+t\left(\overline{\alpha}\right)\langle\langle a_{\overline{\alpha}\sigma} | a_{\alpha\sigma}^{+} \rangle\rangle+t\left(\beta\right)\sum\limits_{k}\langle\langle c_{L\left(R\right)k\sigma} | a_{\alpha\sigma}^{+} \rangle\rangle,\label{eqG3}\\
&&\left(z-\xi_{k\sigma}\right)\langle\langle c_{L\left(R\right)k\sigma} | a_{\alpha\sigma}^{+} \rangle\rangle=t\left(\beta\right)\langle\langle a_{\beta\sigma} | a_{\alpha\sigma}^{+} \rangle\rangle,\label{eqG4}
\end{eqnarray}
where $t\left(\alpha=2\right)\equiv t_{1}$, $t\left(\alpha=3\right)\equiv t_{2}$, $t\left(\beta=1\right)\equiv t_{L}$, $t\left(\beta=4\right)\equiv t_{R}$. In the above equations, besides the first order Green's functions, which we are interested in, the second order Green's functions $\langle\langle n_{\alpha\overline{\sigma}}a_{\alpha\sigma} | a_{\alpha\sigma}^{+} \rangle\rangle$, $\langle\langle n_{\overline{\alpha}\sigma}a_{\alpha\sigma} | a_{\alpha\sigma}^{+} \rangle\rangle$, $\langle\langle n_{\overline{\alpha}\overline{\sigma}}a_{\alpha\sigma} | a_{\alpha\sigma}^{+} \rangle\rangle$, $\langle\langle n_{\beta\overline{\sigma}}a_{\beta\sigma} | a_{\alpha\sigma}^{+} \rangle\rangle$ appear. The equations for them generate third order Green's functions and so on. To receive closed set of equations the decoupling scheme of You and Zheng \cite{you-99a,you-99b,rajput-10} is used. This approximation is valid for temperatures higher than the Kondo temperature \cite{lacroix-81}. In this truncation procedure the intra- and interdot Coulomb correlations are taken into account beyond the Hartree-Fock approximation while spin-flip processes are neglected. Finally, we obtain the following equations,
\begin{eqnarray}
&&\langle\langle a_{\alpha\sigma} | a_{\alpha\sigma}^{+} \rangle\rangle=\left(g_{\alpha\sigma}-K_{\alpha\sigma}\right)\biggl(1+t_{0}\langle\langle a_{\overline{\alpha}\sigma} | a_{\alpha\sigma}^{+} \rangle\rangle+t\left(\alpha\right)\left[\langle\langle a_{\beta\sigma} | a_{\alpha\sigma}^{+} \rangle\rangle+\langle\langle a_{\overline{\beta}\sigma} | a_{\alpha\sigma}^{+} \rangle\rangle \right]\biggr),\nonumber\\
&&\langle\langle a_{\beta\sigma} | a_{\alpha\sigma}^{+} \rangle\rangle=\frac{g_{\beta\sigma}^{\left(0\right)}}{1-\Sigma_{\beta}g_{\beta\sigma}^{\left(0\right)}}
\left[t\left(\alpha\right)\langle\langle a_{\alpha\sigma} | a_{\alpha\sigma}^{+} \rangle\rangle+t\left(\overline{\alpha}\right)\langle\langle a_{\overline{\alpha}\sigma} | a_{\alpha\sigma}^{+} \rangle\rangle \right],\label{eqG5}
\end{eqnarray}
where
\begin{eqnarray}
&&K_{\alpha\sigma}=\frac{UV\langle a_{\alpha\overline{\sigma}}^{+}a_{\overline{\alpha}\overline{\sigma}}\rangle^2}
{b_{\alpha1}b_{\alpha\sigma4}}\left(\frac{1}{b_{\alpha2}}+\frac{1}{b_{\alpha\sigma3}}\right),\nonumber\\
&&g_{\alpha\sigma}=g_{\alpha\sigma}^{\left(0\right)}+\frac{V}{b_{\alpha1}}\left[g_{\alpha\sigma}^{\left(1\right)}+
g_{\alpha\sigma}^{\left(2\right)}+\frac{U}{b_{\alpha2}}g_{\alpha\sigma}^{\left(3\right)}\right],~g_{\alpha\sigma}^{\left(0\right)}=\frac{1-\langle n_{\alpha\overline{\sigma}}\rangle}{b_{\alpha1}}+\frac{\langle n_{\alpha\overline{\sigma}}\rangle}{b_{\alpha2}},\label{eqG6}\\
&&g_{\alpha\sigma}^{\left(1\right)}=\frac{\left(1-\langle n_{\alpha\overline{\sigma}}\rangle\right)\langle n_{\overline{\alpha}\sigma}\rangle}{b_{\alpha\overline{\sigma}3}}+\frac{\langle n_{\alpha\overline{\sigma}}\rangle\langle n_{\overline{\alpha}\sigma}\rangle}{b_{\alpha\overline{\sigma}4}},~g_{\alpha\sigma}^{\left(2\right)}=\frac{\left(1-\langle n_{\alpha\overline{\sigma}}\rangle\right)\langle n_{\overline{\alpha}\overline{\sigma}}\rangle}{b_{\alpha\sigma3}}+\frac{\langle n_{\alpha\overline{\sigma}}\rangle\langle n_{\overline{\alpha}\overline{\sigma}}\rangle}{b_{\alpha\sigma4}},\nonumber\\
&&g_{\alpha\sigma}^{\left(3\right)}=\frac{\langle n_{\alpha\overline{\sigma}}\rangle\langle n_{\overline{\alpha}\sigma}\rangle}{b_{\alpha\overline{\sigma}4}}+\frac{\langle n_{\alpha\overline{\sigma}}\rangle\langle n_{\overline{\alpha}\overline{\sigma}}\rangle}{b_{\alpha\sigma4}},~g_{\beta\sigma}^{\left(0\right)}=g_{\alpha \rightarrow \beta,\sigma}^{\left(0\right)},~\Sigma_{\beta}=-i\frac{t}{2},\nonumber\\
&&b_{\alpha1}=z-\xi_{\alpha},~b_{\alpha2}=b_{\alpha1}-U,~b_{\alpha\sigma3}=b_{\alpha1}-V\left(1+\langle
n_{\overline{\alpha}\sigma}\rangle\right) ,~b_{\alpha\sigma4}=b_{\alpha\sigma3}-U.\nonumber
\end{eqnarray}
Solving the system \eqref{eqG5} and using the notations of \cite{rajput-10} we get in non-magnetic case, $\langle n_{i\sigma}\rangle=\langle n_{i\overline{\sigma}}\rangle$, $\langle a_{i\sigma}^{+}a_{j\sigma}\rangle=\langle a_{i\overline{\sigma}}^{+}a_{j\overline{\sigma}}\rangle$, the following expressions for the matrix elements of $\hat{G}^{r}$,
\begin{eqnarray}
&&G^{r}_{\alpha\alpha}=\frac{C_{\alpha}\bigl(D_{\overline{\alpha}}D_{\beta}-2t^2\left(\overline{\alpha}\right)C_{\overline{\alpha}}C_{\beta}+\frac{i}{2}\Gamma C_{\beta}D_{\overline{\alpha}}\bigr)}{X_{1}-2X_{2}+iY},\nonumber\\
&&G^{r}_{\alpha\overline{\alpha}}=\frac{C_{\alpha}C_{\overline{\alpha}}\left(t_{0}D_{\beta}+2t\left(\alpha\right)t\left(\overline{\alpha}\right)C_{\beta}+\frac{i}{2}\Gamma t_{0}C_{\beta}\right)}{X_{1}-2X_{2}+iY},\nonumber\\
&&G^{r}_{\alpha\beta}=\frac{C_{\alpha}C_{\beta}\left(t\left(\alpha\right)D_{\overline{\alpha}}+t_{0}t\left(\overline{\alpha}\right)C_{\overline{\alpha}}\right)}{X_{1}-2X_{2}+iY},\label{GF}\\
&&G^{r}_{\beta\beta}=\frac{C_{\beta}\left(X_{1}-X_{2}+iY\right)}{\left(D_{\beta}+\frac{i}{2}\Gamma C_{\beta}\right)\left(X_{1}-2X_{2}+iY\right)},\nonumber\\
&&G^{r}_{\overline{\beta}\beta}=\frac{C_{\beta}X_{2}}{\left(D_{\beta}+\frac{i}{2}\Gamma C_{\beta}\right)\left(X_{1}-2X_{2}+iY\right)},\nonumber
\end{eqnarray}
where
\begin{eqnarray}
&&D_{\alpha}=b_{\alpha1}b_{\alpha2}b_{\alpha\sigma3}b_{\alpha\sigma4},~D_{\beta}=b_{\beta1}b_{\beta2},\nonumber\\
&&C_{\beta}=b_{\beta2}+U\langle n_{\beta\sigma}\rangle,~C_{\alpha}=C_{\alpha1}+C_{\alpha2},\nonumber\\
&&C_{\alpha1}=b_{\alpha\sigma4}\left(b_{\alpha2}b_{\alpha\sigma3}+Ub_{\alpha\sigma3}\langle n_{\alpha\sigma}\rangle+2Vb_{\alpha2}\langle n_{\overline{\alpha}\sigma}\rangle\right),\nonumber\\
&&C_{\alpha2}=UV\left(b_{\alpha2}+b_{\alpha\sigma3}\right)\left(2\langle n_{\alpha\sigma}\rangle\langle n_{\overline{\alpha}\sigma}\rangle-\langle a_{\alpha\sigma}^{+}a_{\overline{\alpha}\sigma}\rangle^2\right),\\
&&X_{1}=D_{\beta}\left(D_{\alpha}D_{\overline{\alpha}}-t_{0}^2C_{\alpha}C_{\overline{\alpha}}\right),\nonumber\\
&&X_{2}=C_{\beta}\left[t^2\left(\alpha\right)C_{\alpha}D_{\overline{\alpha}}+
t^2\left(\overline{\alpha}\right)C_{\overline{\alpha}}D_{\alpha}+2t_{0}t\left(\alpha\right)t\left(\overline{\alpha}\right)C_{\alpha}C_{\overline{\alpha}}\right],\nonumber\\
&&Y=\frac{1}{2}\Gamma C_{\beta}\left(D_{\alpha}D_{\overline{\alpha}}-t_{0}^2C_{\alpha}C_{\overline{\alpha}}\right).\nonumber
\end{eqnarray}
The occupation numbers and correlators are found self-consistently using the kinetic equations in equilibrium:
\begin{eqnarray}
&&\langle n_{i\sigma}\rangle=-\frac{1}{\pi}\int d\omega f\left(\omega\right) Im\left[G^{r}_{ii}\left(\omega\right)\right],\nonumber\\
&&\langle a_{i\sigma}^{+}a_{j\sigma}\rangle=-\frac{1}{\pi}\int d\omega f\left(\omega\right) Im\left[G^{r}_{ji}\left(\omega\right)\right],~i,~j=\alpha,~\beta. \label{kineq}
\end{eqnarray}

For the sake of simplicity we will analyze a symmetrical transport situation and use $\Gamma_{L}=\Gamma_{R}=t$ in energy units. In this article we consider the strong coupling regime, $t=t_{1}$.

\begin{center}\textbf{5. Transport properties of the QQD device without the Coulomb interactions}\end{center}
\begin{center}\textbf{A. Isotropic QQD}\end{center}

We start the analysis with the case when the couplings between the left (right) QD and both QDs in the middle part are the same, $t_1=t_2=1$ \cite{yan-13}, and temperature close to zero, $k_{B}T=10^{-6}~t$.
\begin{figure}[h!] \begin{center}
\includegraphics[width=0.48\textwidth]{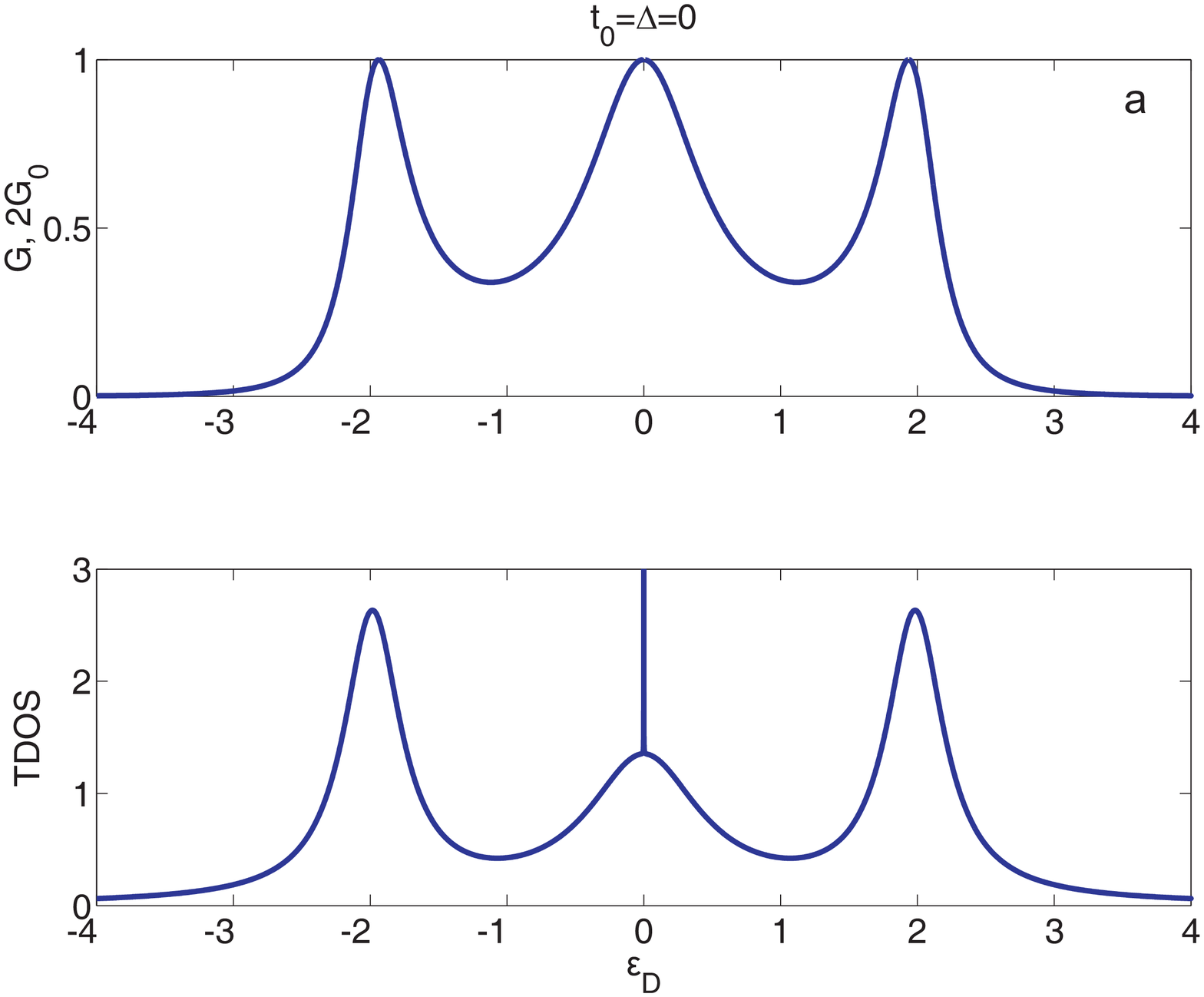}
\includegraphics[width=0.48\textwidth]{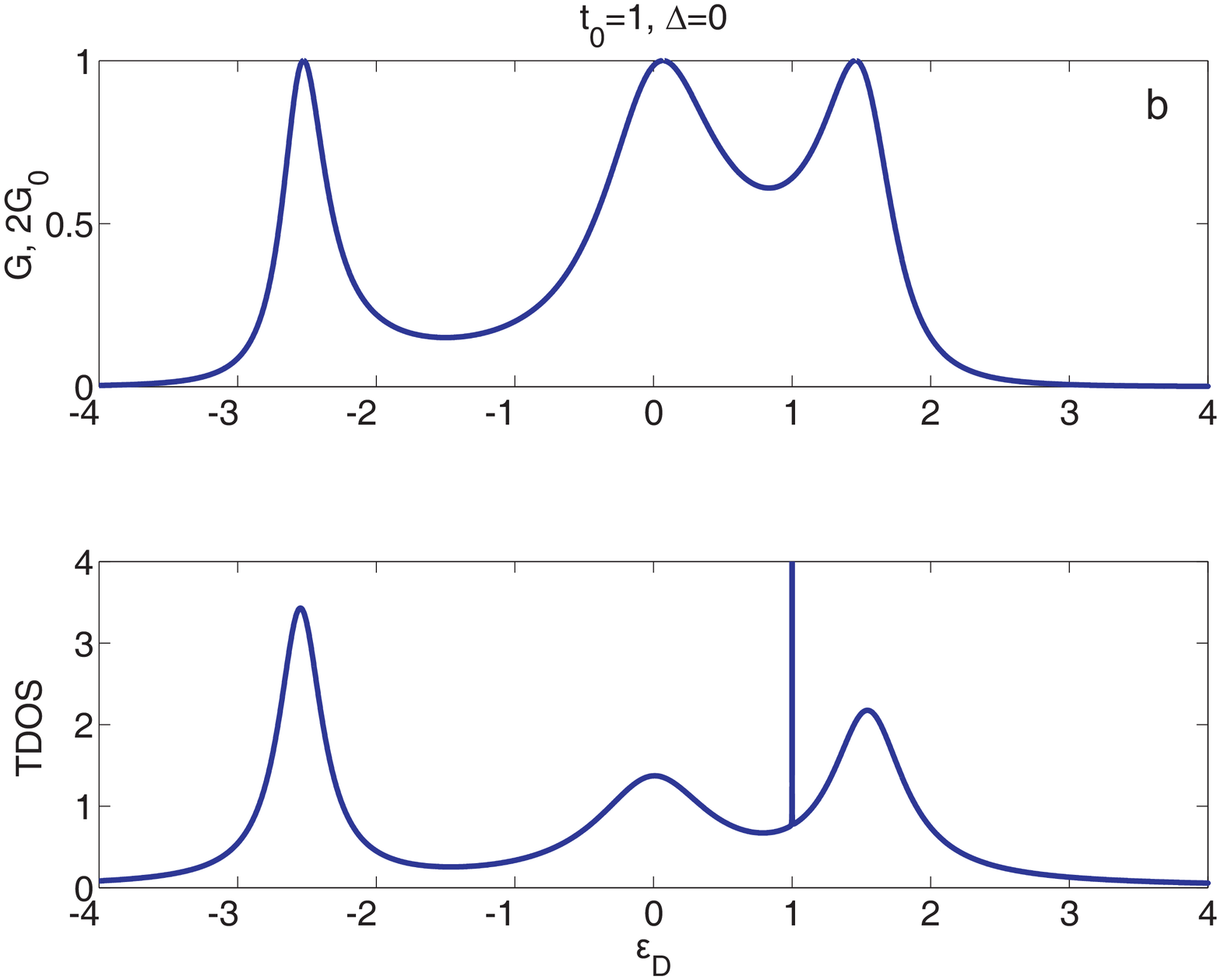}
\includegraphics[width=0.48\textwidth]{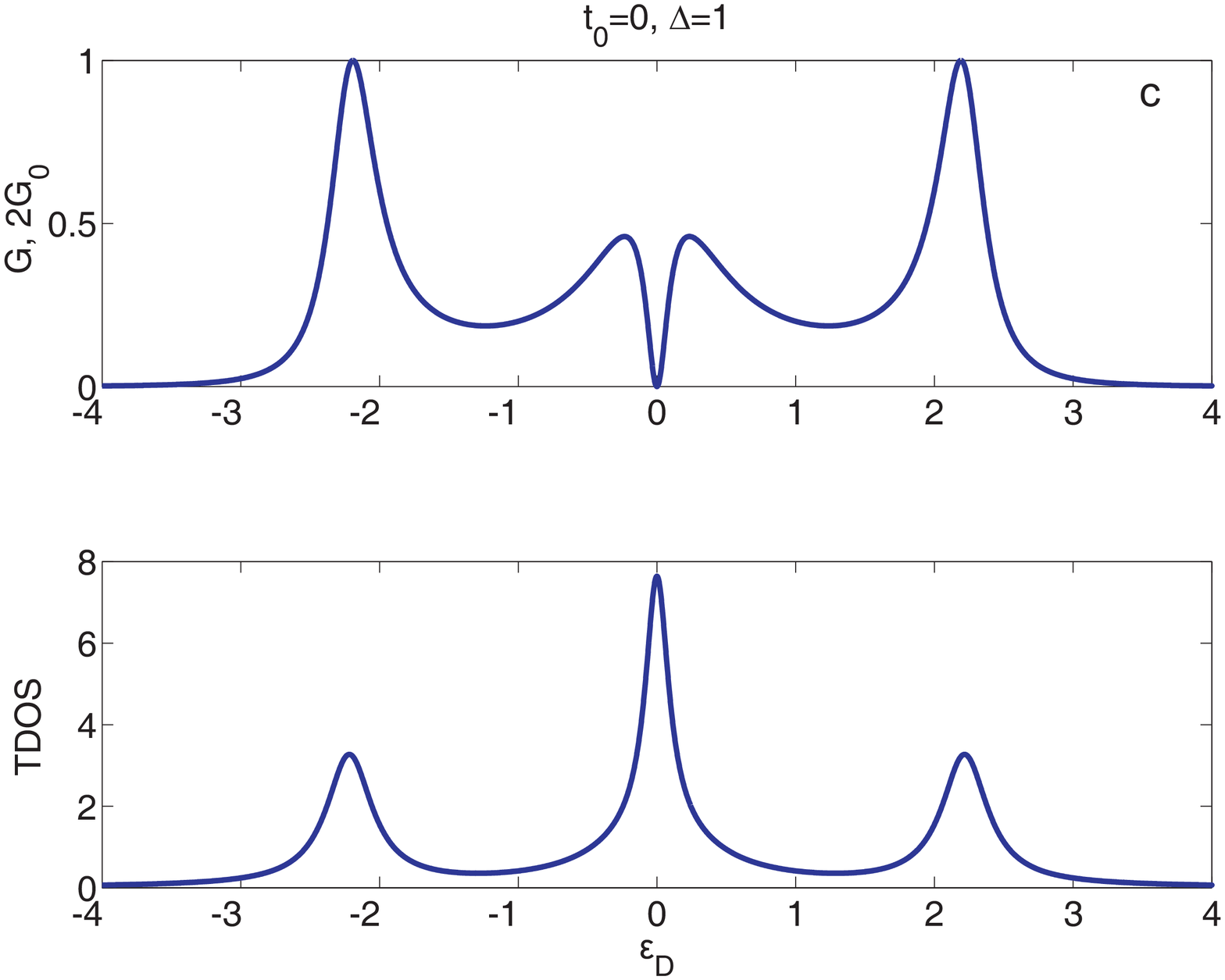}
\includegraphics[width=0.48\textwidth]{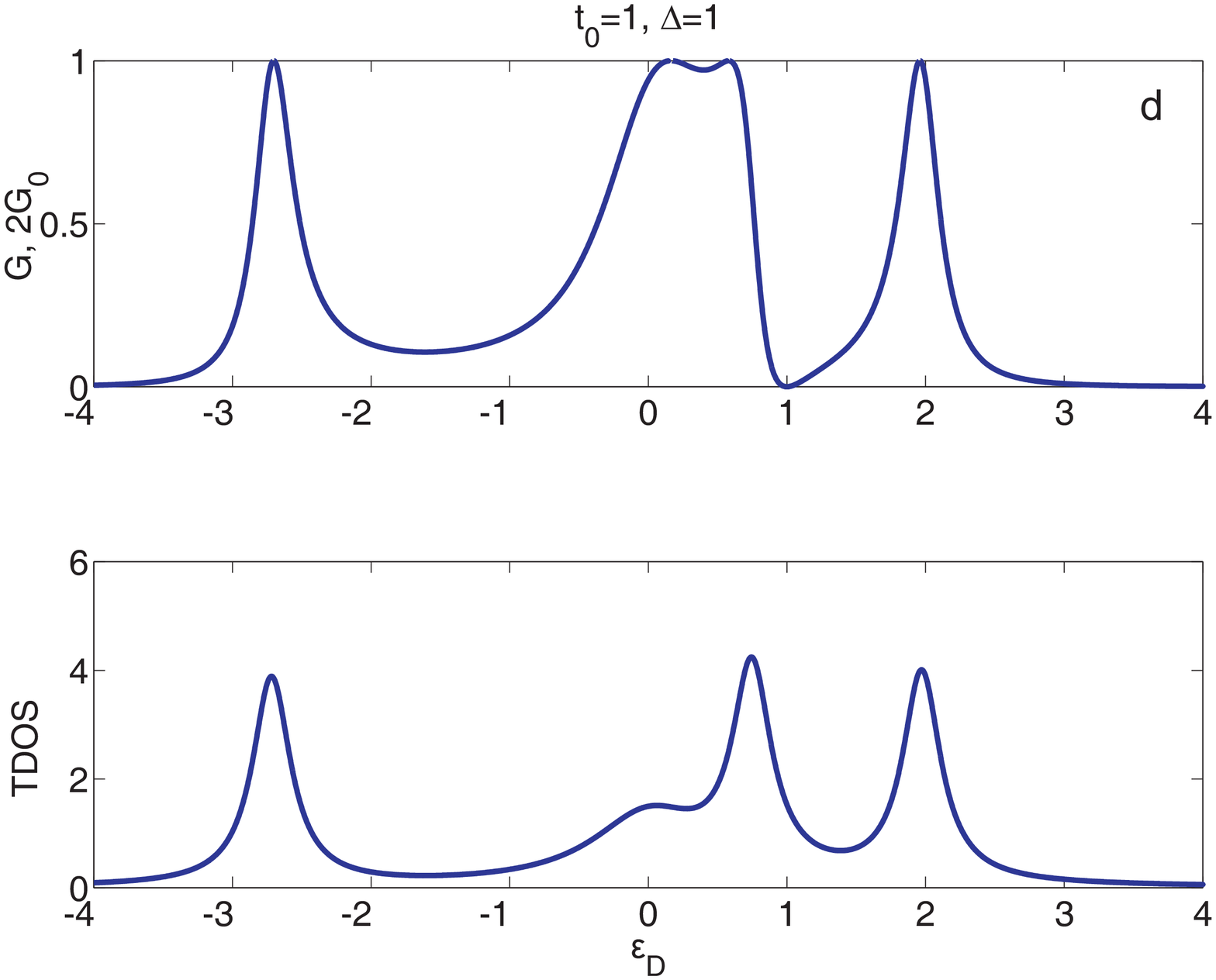}
\caption{The conductance and TDOS of the isotropic QQD: a) $t_0=\Delta=0$; b) $t_0=1,~\Delta=0$; c) $t_0=0,~\Delta=1$; d) $t_0=1,~\Delta=1$.} \label{2}
\end{center}
\end{figure}

The simplest transport situation occurs if all levels have the same energy, $\xi_{1\sigma}=\xi_{2\sigma}=\xi_{3\sigma}=\xi_{4\sigma}$, and $t_{0}=0$. The function $G\left(\varepsilon_D\right)$ is depicted at figure \ref{2}a. The triple-peak structure (TPS) of the conductance can be easily understood since the system can be treated as the one consists of two arms each composed of three coupled QDs. The corresponding TDOS has maxima at the same positions of $\varepsilon_D$. Additionally, the bound state in continuum (BIC) appears at $\varepsilon_D=0$ \cite{volya-03,sadreev-06} - the sharp peak with nearly zero width due to $i\delta$ term in $\widehat{G}^{r}$ \eqref{Gr}. The position of the BIC depends on $t_0$. In particular it shifts toward higher energies when $t_0$ increases (see TDOS at fig. \ref{2}b). Simultaneously the conductance spectrum doesn't contain corresponding features that is exactly BIC's property. In contrast to \cite{yan-13} we show that there are more than one way to make finite lifetime of this state for such a system. First of them is to realize two nonequivalent transport channels by means of the energy shift $\Delta$, $\xi_{2\sigma}=\xi_{3\sigma}+2\Delta$ \cite{yan-13}. As a result destructive interference of the electronic waves propagating along these two paths gives rise to the Fano antiresonance \cite{fano-61} in the conductance spectrum and the resonance with finite width in the TDOS at $\varepsilon_D=0$ (see fig. \ref{2}c). If $t_{0}\neq0$ the Fano antiresonance transforms to the Fano-Feshbach asymmetrical peak in the conductance spectrum at figure \ref{2}d.

\begin{center}\textbf{B. Anisotropic QQD}\end{center}

\begin{figure}[h!] \begin{center}
\includegraphics[width=0.48\textwidth]{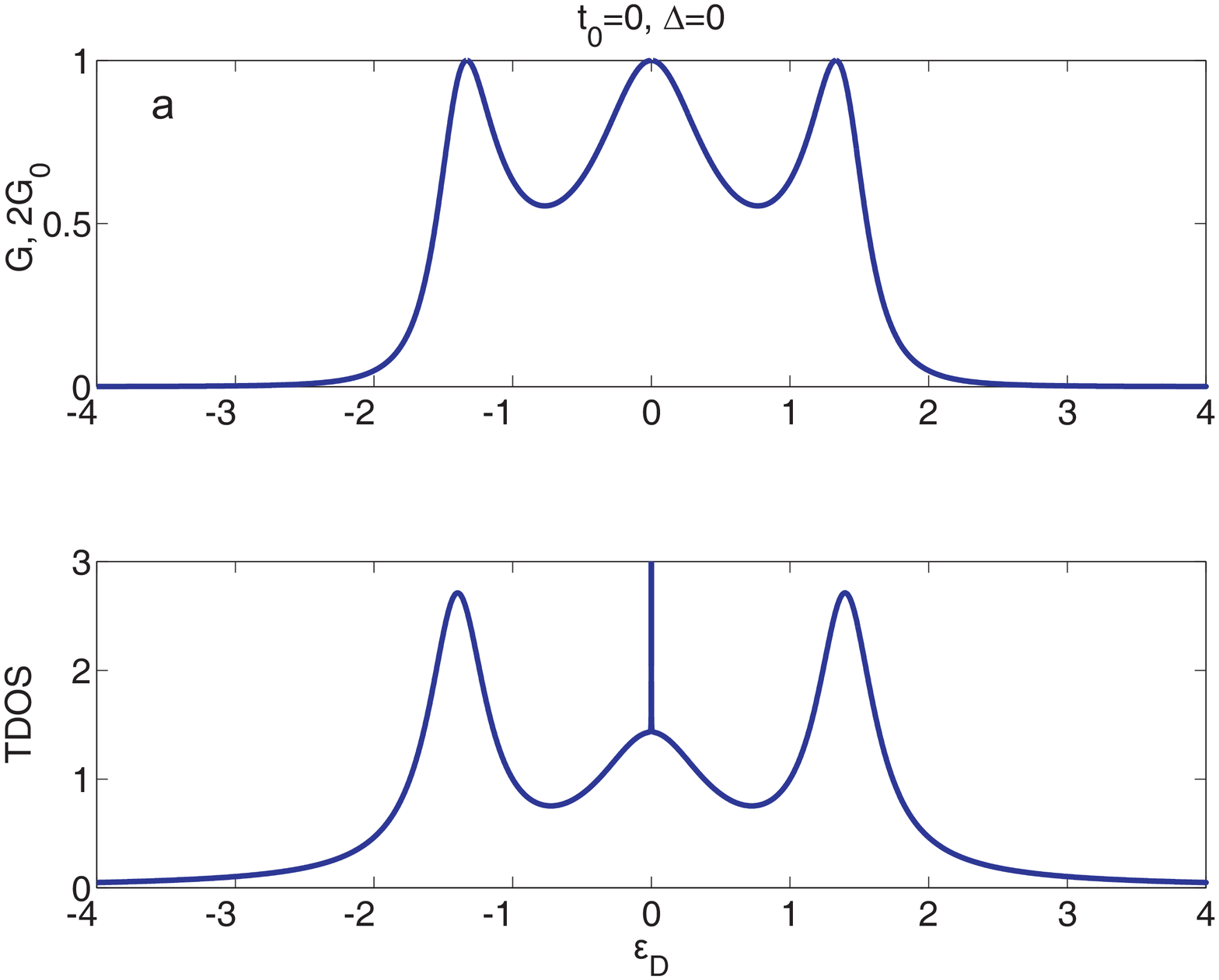}
\includegraphics[width=0.48\textwidth]{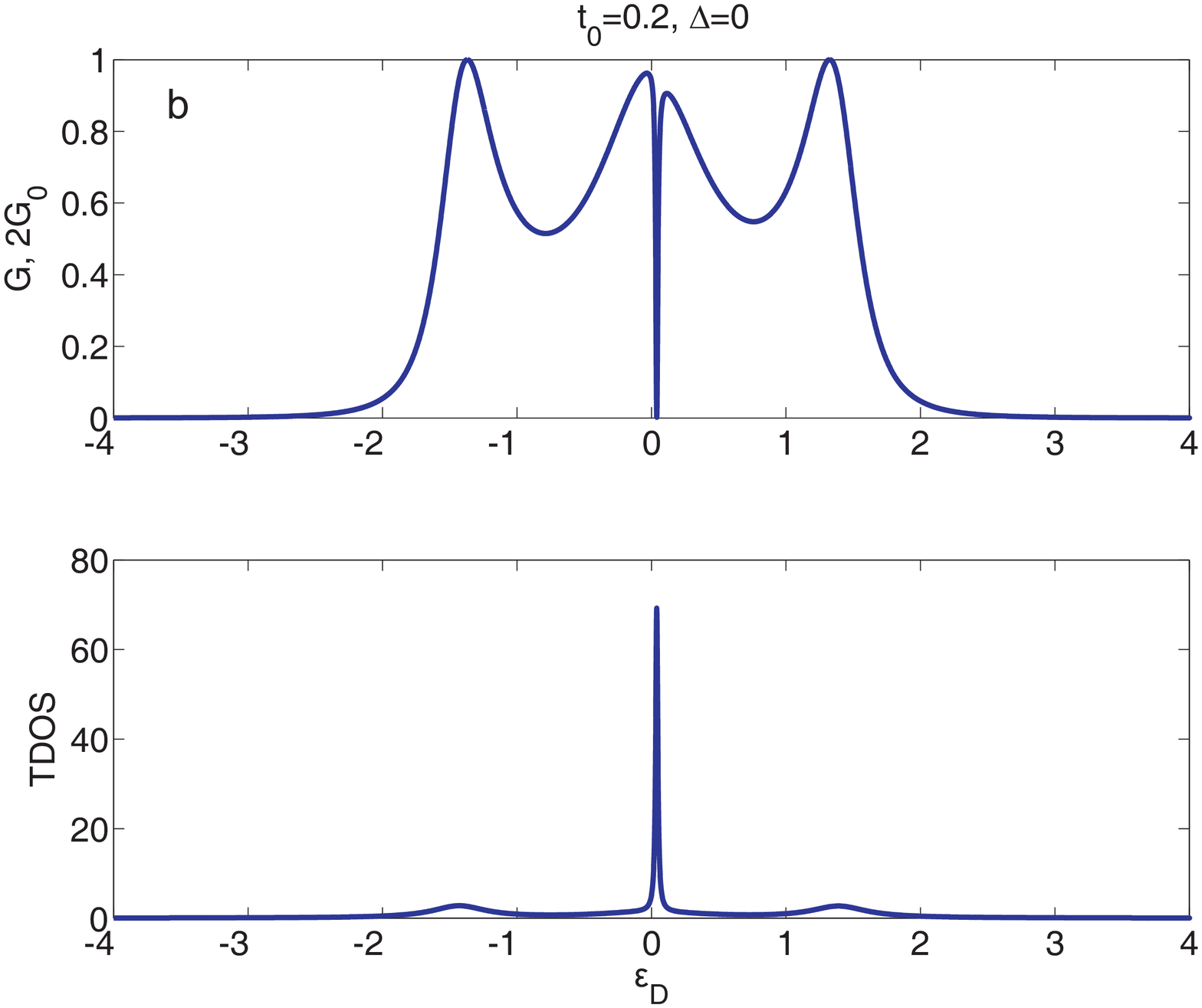}
\includegraphics[width=0.48\textwidth]{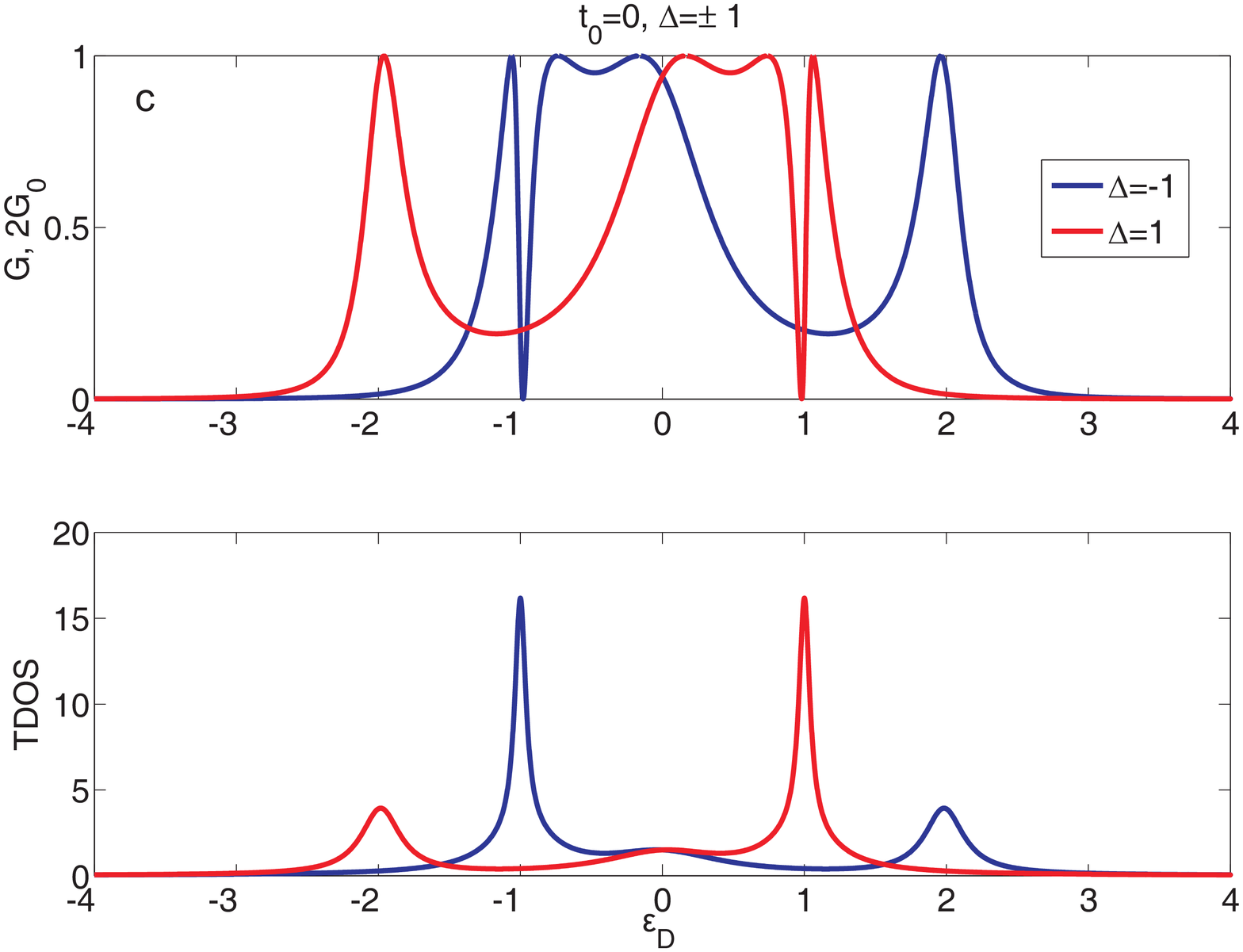}
\includegraphics[width=0.48\textwidth]{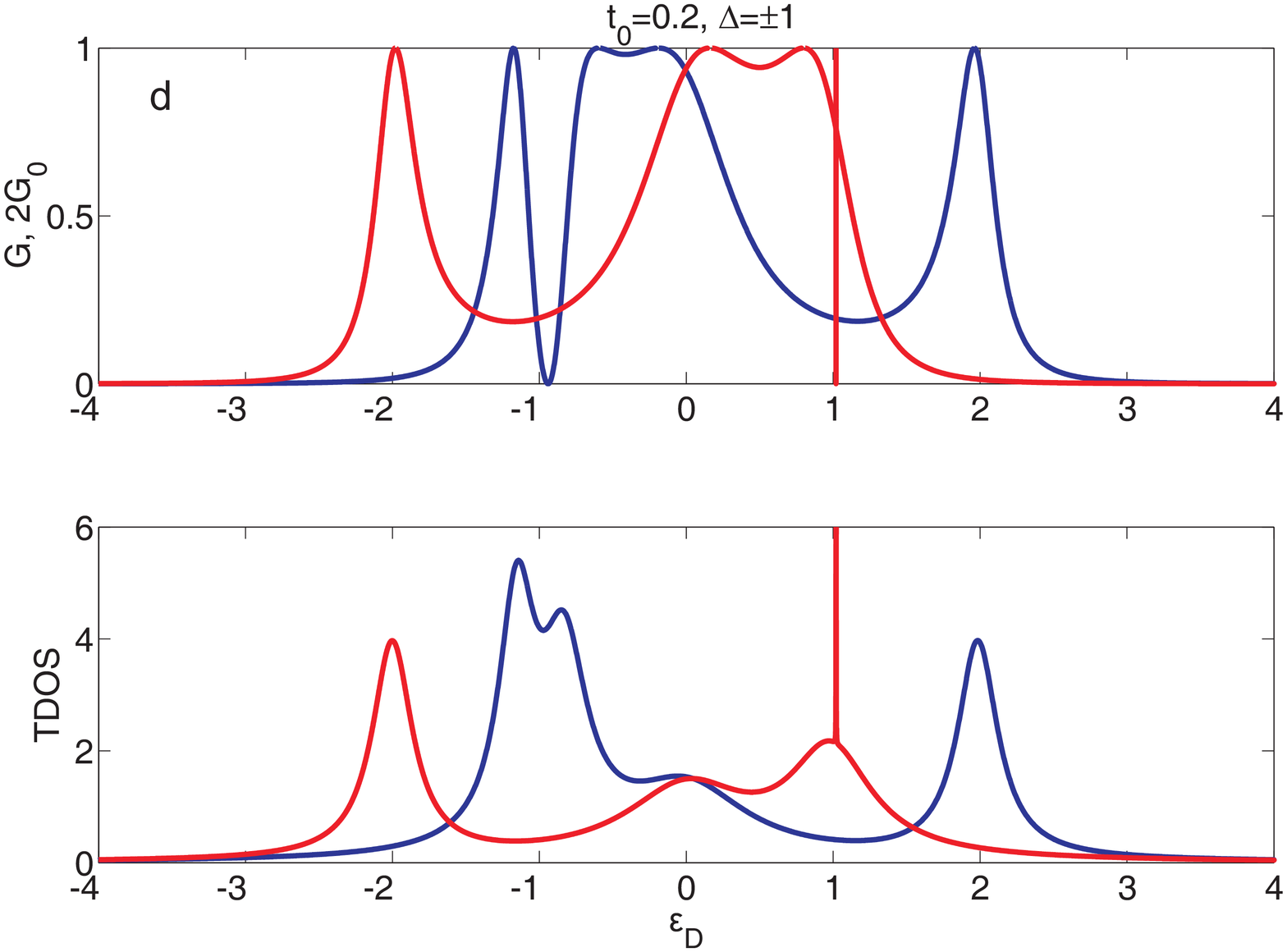}
\includegraphics[width=0.48\textwidth]{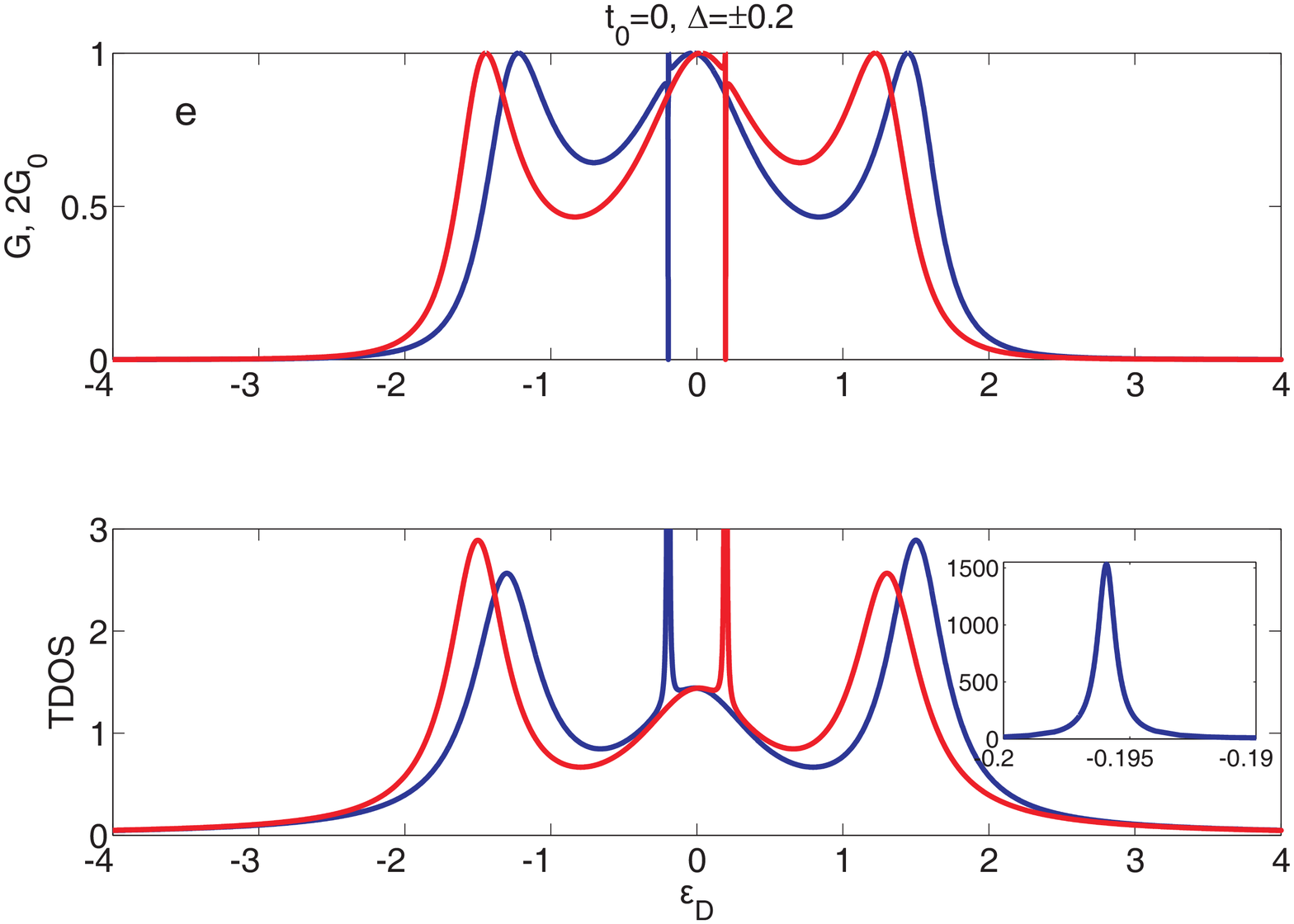}
\includegraphics[width=0.48\textwidth]{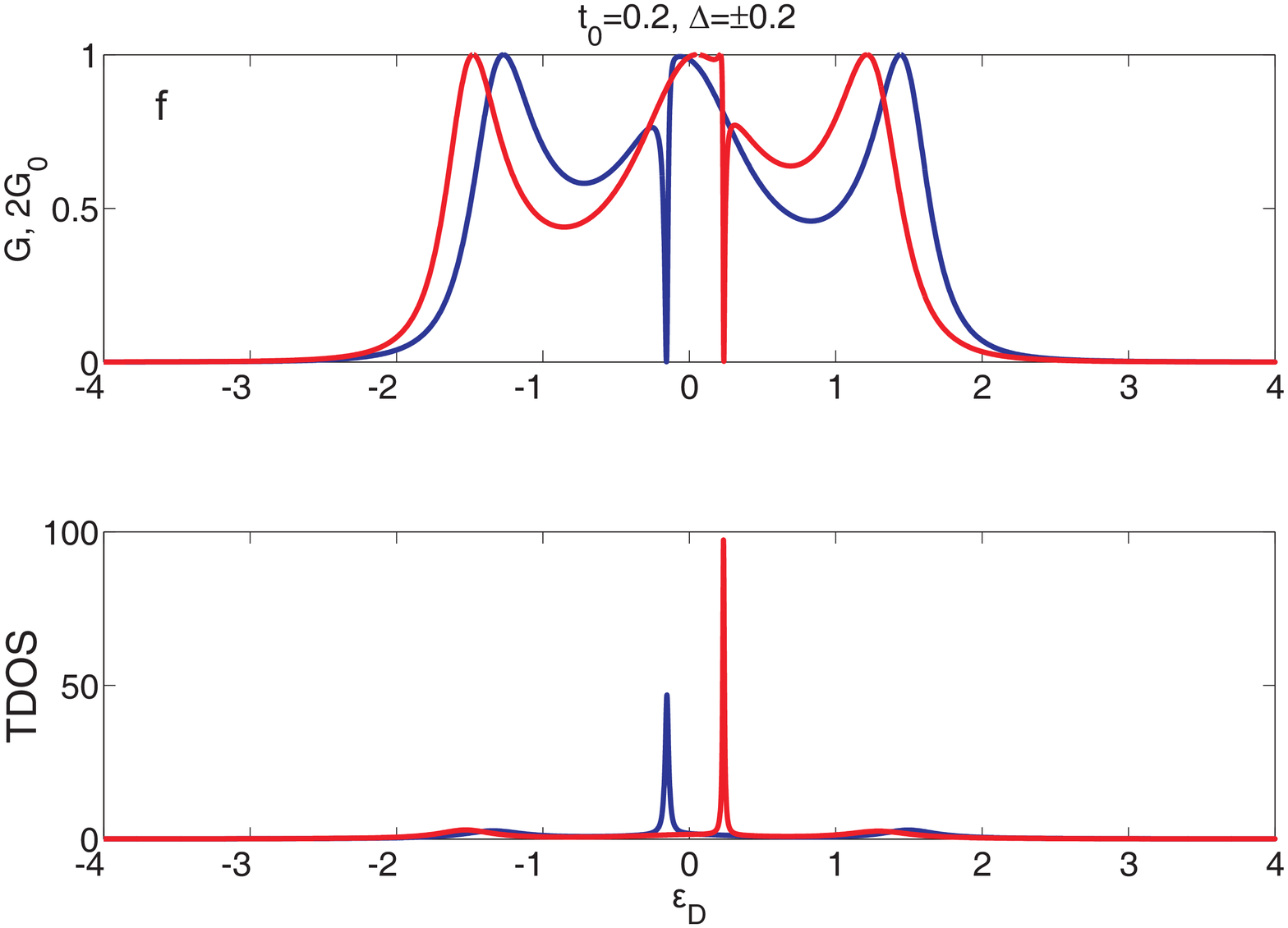}
\caption{The conductance and TDOS of the anisotropic QQD, $t_{2}=0.1$: a) $t_0=\Delta=0$; b) $t_0=0.2,~\Delta=0$; c) $t_0=0,~\Delta=\pm1$; d) $t_0=0.2,~\Delta=\pm1$; e) $t_0=0,~\Delta=\pm0.2$; f) $t_0=0.2,~\Delta=\pm0.2$.} \label{3}
\end{center}
\end{figure}
Let us consider the transport regime where the couplings between the QDs, $t_1,~t_2,~t_0$, are different and $t_1 \gg t_2,~t_0$. Such an anisotropy is more convenient for real systems and the inequality between parameters can be even enhanced by the EPE. Numerical calculations show that the anisotropy is the new mechanism leading to the Fano-Feshbach resonance along with the above-described one \cite{yan-13}. The simple explanation of this effect is based on the interpretation of the 2nd and 3rd QDs as an artificial molecule, a dimer \cite{lu-05,liu-07}. The dimer has bonding and antibonding eigenstates which in general coupled to other part of the system unequally. Then more broadened level is treated as a continuum or nonresonant channel whereas less broadened level plays the role of a discrete level or resonant channel in the original Fano picture \cite{fano-61}. The phase of the wave function in the nonresonant channel changes slightly as the energy passes an interval $\sim \Gamma$, where $\Gamma$ is the broadening of the discrete level. However, the phase in the resonant channel shifts by $\sim \pi$ at the same energy interval. Consequently, the Fano-Feshbach resonant asymmetrical peak appears as a result of constructive and destructive interference at this energy range around the discrete level. Following \cite{lu-05} in the case of $\Delta=0$, the coupling with one of the molecular states is absent without the anisotropy, $t_1=t_2$, and there is no the Fano-Feshbach effect (see fig. \ref{2}b). In the opposite case the anisotropy induces the corresponding antiresonance as it is depicted at figure \ref{3}b. If $\Delta\neq0$ the coupling with both bonding and antibonding states doesn't equal to zero even though $t_1=t_2$ and the Fano-Feshbach resonance occurs (see fig. \ref{2}c, d) \cite{liu-07}. The anisotropy in this case leads to the change of the shape and width of the resonance and its dependence on the sign of $\Delta$ (see red and blue curves at fig. \ref{3}c-f for $\Delta=\pm1$ respectively). Specifically, the significant difference is observed at figures \ref{3}d where the wide resonance corresponds to $\xi_{2\sigma}<\xi_{3\sigma}$ and very narrow one to $\xi_{2\sigma}>\xi_{3\sigma}$. The symmetry of the Fano-Feshbach resonance position is broken if $t_0\neq0$.

\begin{center}\textbf{C. Temperature effects}\end{center}

\begin{figure}[h!] \begin{center}
\includegraphics[width=0.45\textwidth]{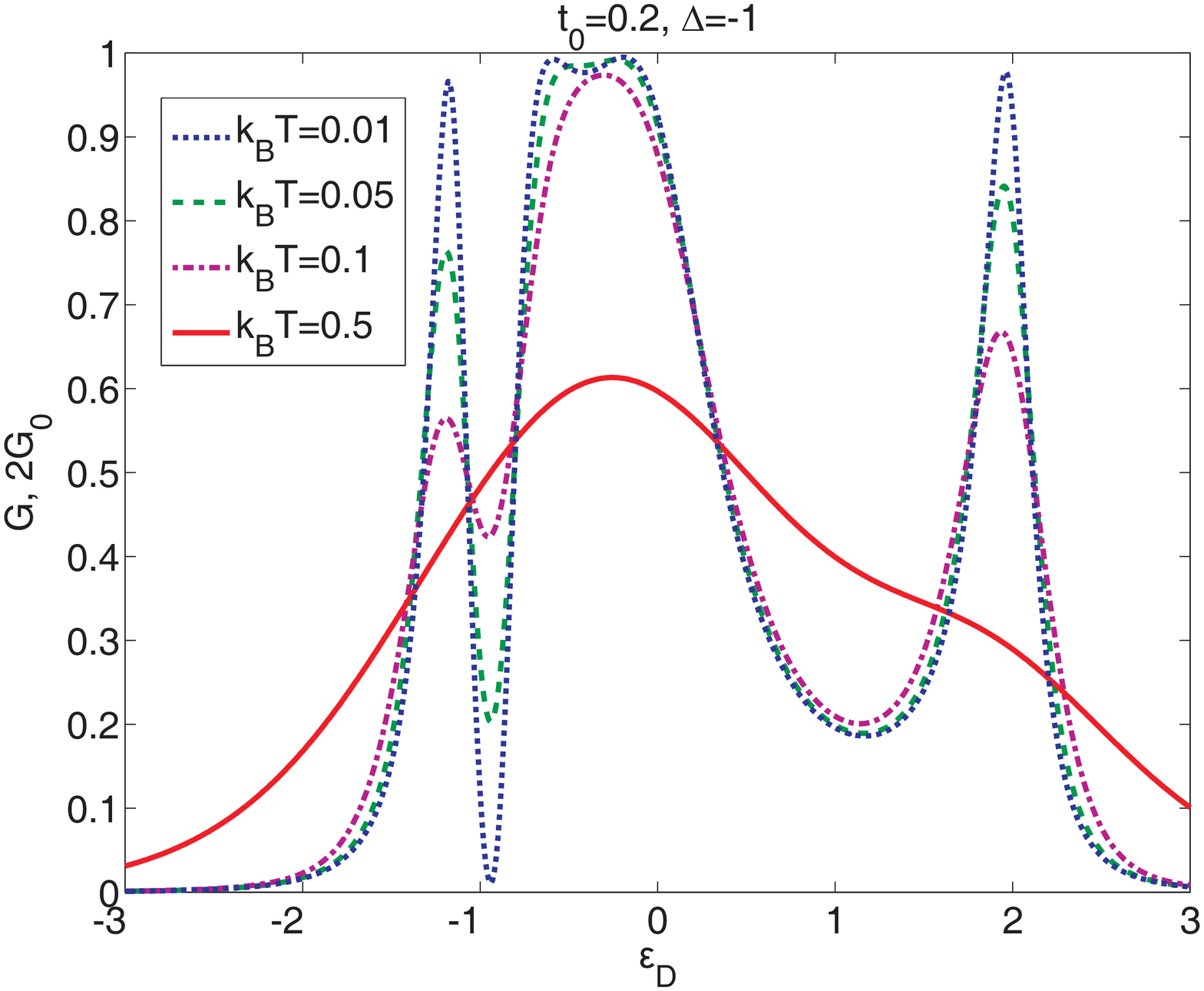}
\caption{The temperature effects on the conductance of the anisotropic QQD structure $t_{2}=0.1$, $t_{0}=0.2$, $\Delta=-1$.}\label{4}
\end{center}
\end{figure}
If the temperature is comparable with the spacing between energy levels of the structure, i.e. $k_{B}T \sim \Delta,~t_{2},~t_{0}$, that the conductance can be calculated as
\begin{equation} \label{ILT}
G=-G_{0}\int\limits_{-\infty}^{+\infty}d\omega Tr\Bigl[\widehat{T}\left(\omega\right)\Bigr] \frac{\partial f}{\partial \omega},
\end{equation}
The temperature influence on the conductance is depicted at figure \ref{4}. If $k_{B}T \ll \Delta,~t_{2},~t_{0}$ that the smearing of the conductance, for example, the Fano-Feshbach asymmetrical peak, isn't strong ($k_{B}T=0.01$, dotted line). The dependence gradually becomes the Lorentzian-like curve with increasing $k_{B}T$.

\begin{center}\textbf{6. The effects of the Coulomb interactions}\end{center}
\begin{center}\textbf{A. Isotropic QQD}\end{center}

\begin{figure}[h!] \begin{center}
\includegraphics[width=0.45\textwidth]{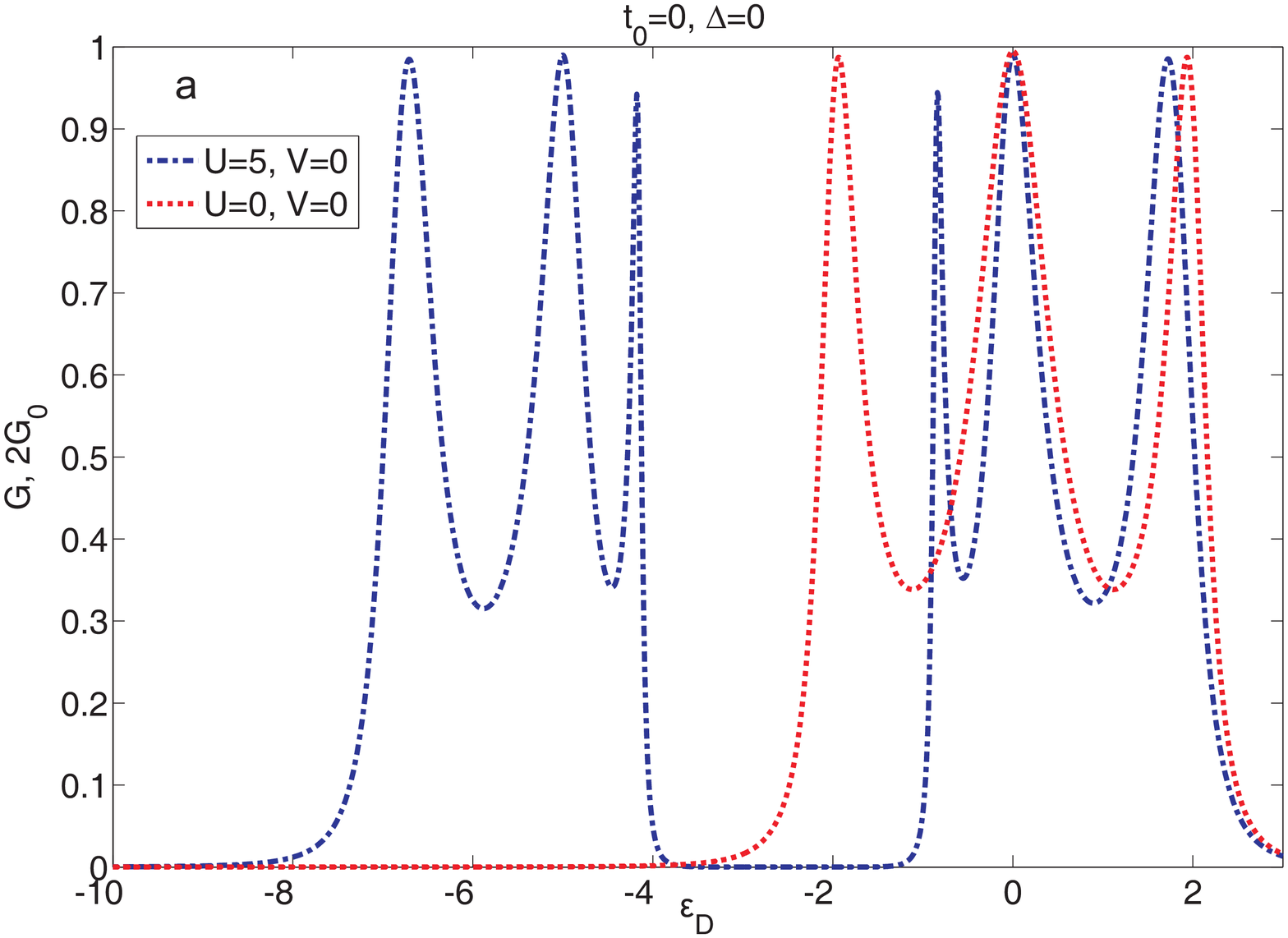}
\includegraphics[width=0.45\textwidth]{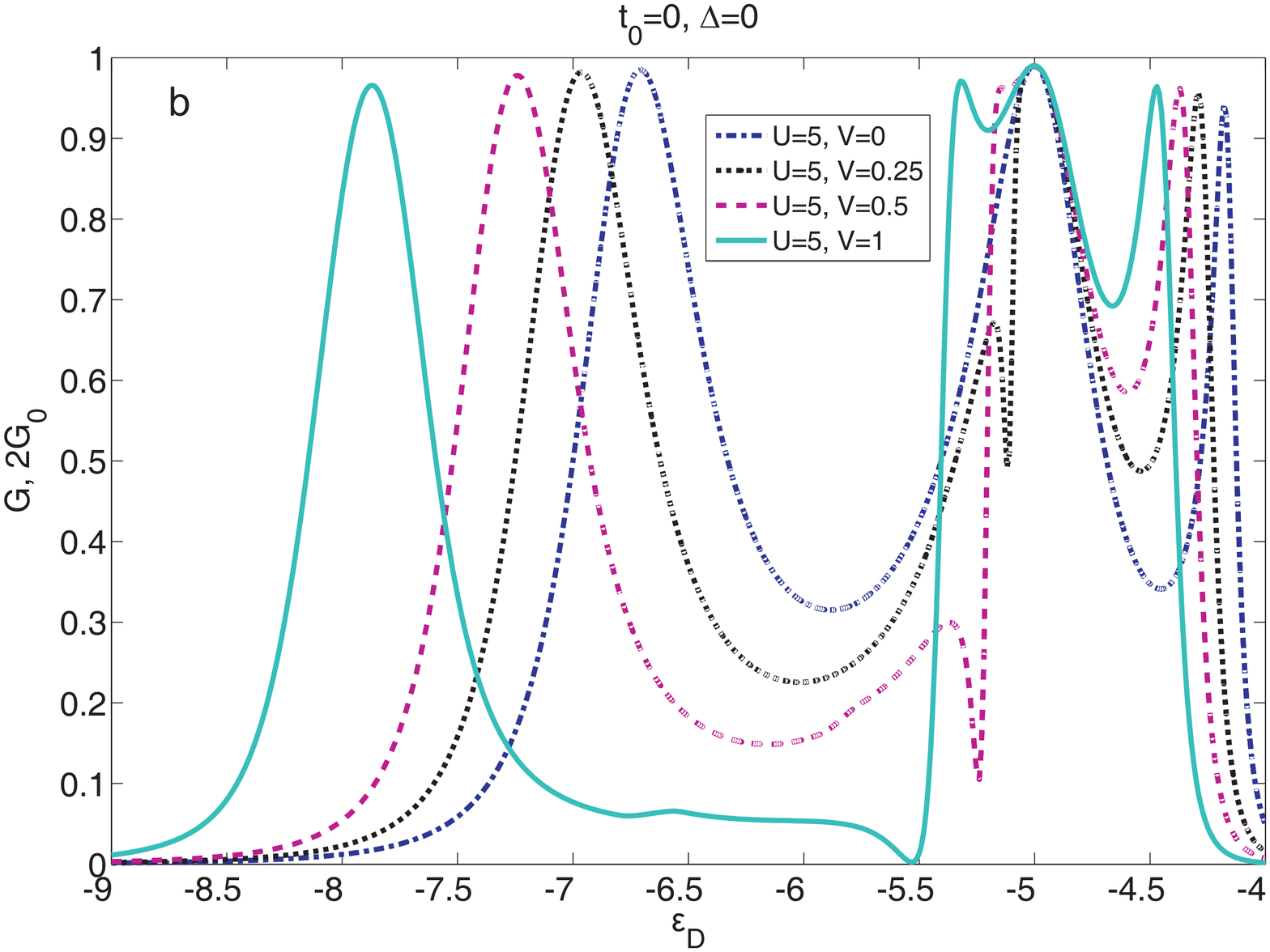}
\caption{The influence of the Coulomb interactions in the QQD on the conductance: a) the effect of the intradot interactions, $U$; b) the effect of the interdot interaction, $V$, $k_{B}T=0.01$.} \label{5}
\end{center}
\end{figure}
The effect of the intradot, $U$, and interdot, $V$, Coulomb interactions on the conductance spectrum is displayed in details at figures \ref{5}a and b respectively. The strong Coulomb repulsion of the electrons with different spin projections in each QD gives rise to the splitting of the TPS (dotted line at fig. \ref{5}a) and the appearance of the well-defined insulating band between two TPSs where $G$ is close to zero (dash-dot line at fig. \ref{5}a). It is clearly seen that the band forms without making the energy difference, $\Delta$, which is influenced by external gate fields, and requires lesser quantity of QDs in comparison with \cite{gong-06,fu-12}. Taking into account the interdot Coulomb interaction of the electrons in the middle part results in the splitting of the central peak in both TPSs as it is depicted at figure \ref{5}b by the example of the left TPS. It is worth noticing that the increase of $V$ gives rise to the Fano antiresonance and the appearance of the sufficiently wide band with low conductance ($G\sim0.1$ at $\varepsilon_D=-7$ --- $-5.5$).
\begin{figure}[h!] \begin{center}
\includegraphics[width=0.45\textwidth]{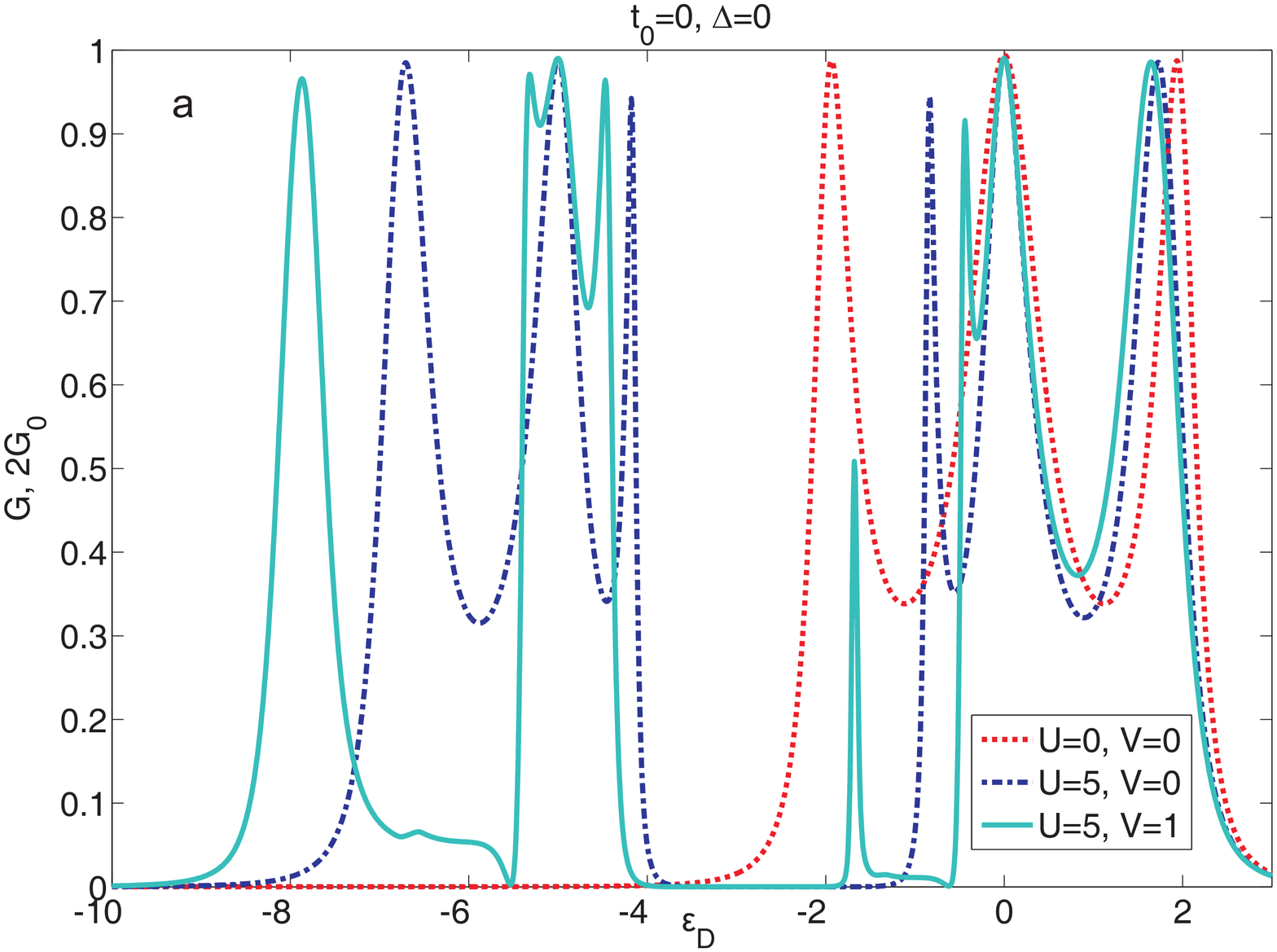}
\includegraphics[width=0.45\textwidth]{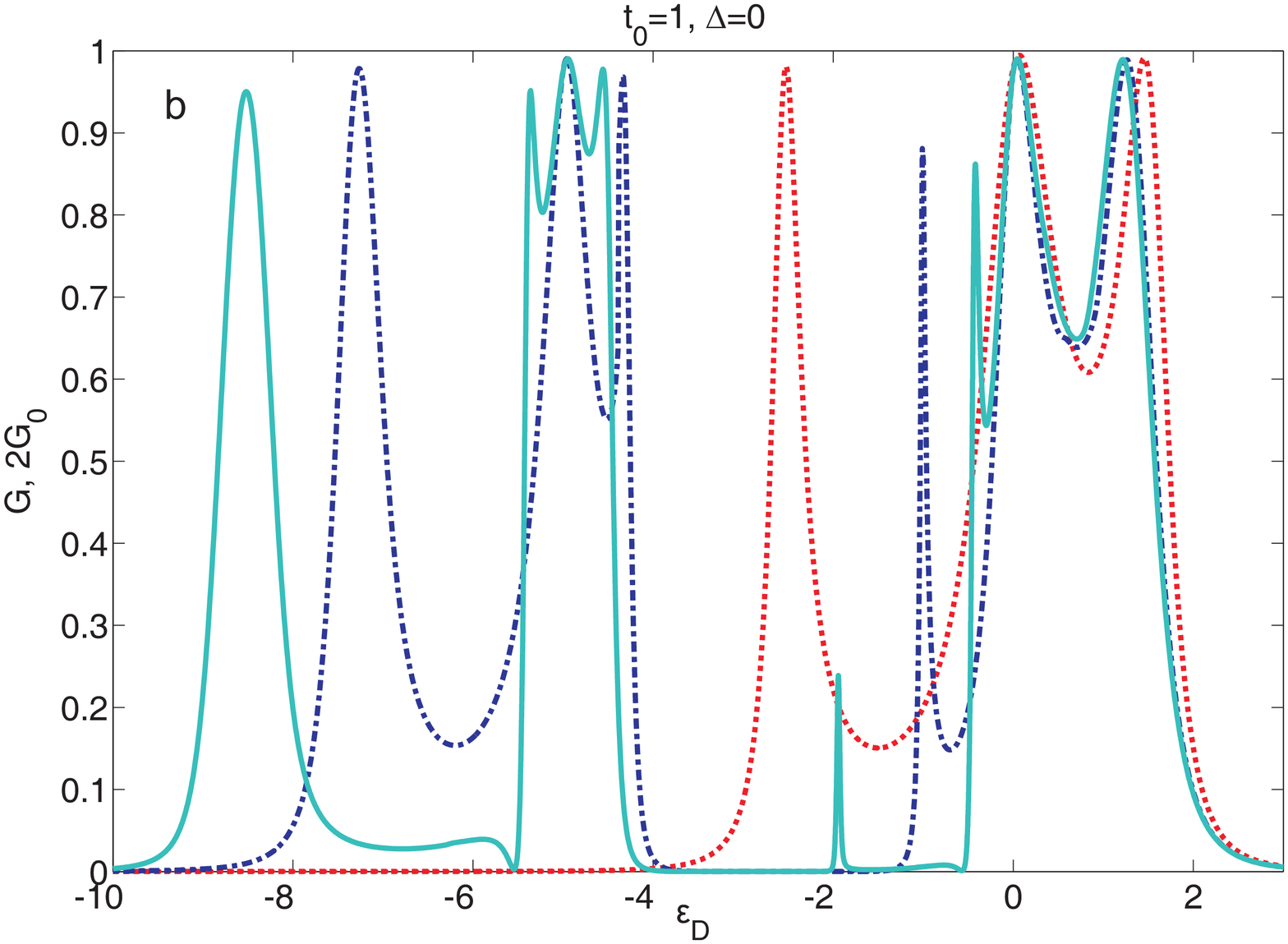}
\includegraphics[width=0.45\textwidth]{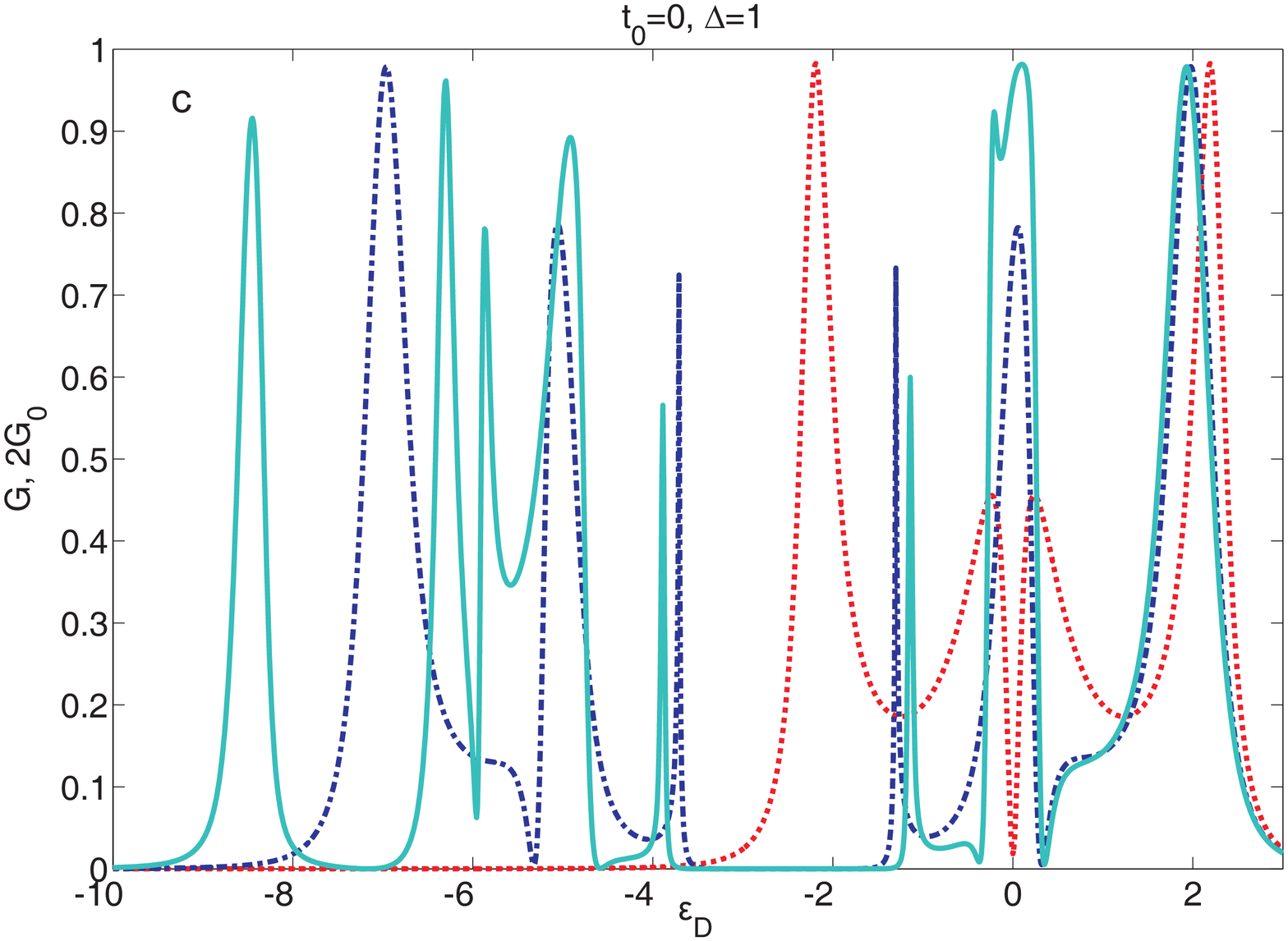}
\includegraphics[width=0.45\textwidth]{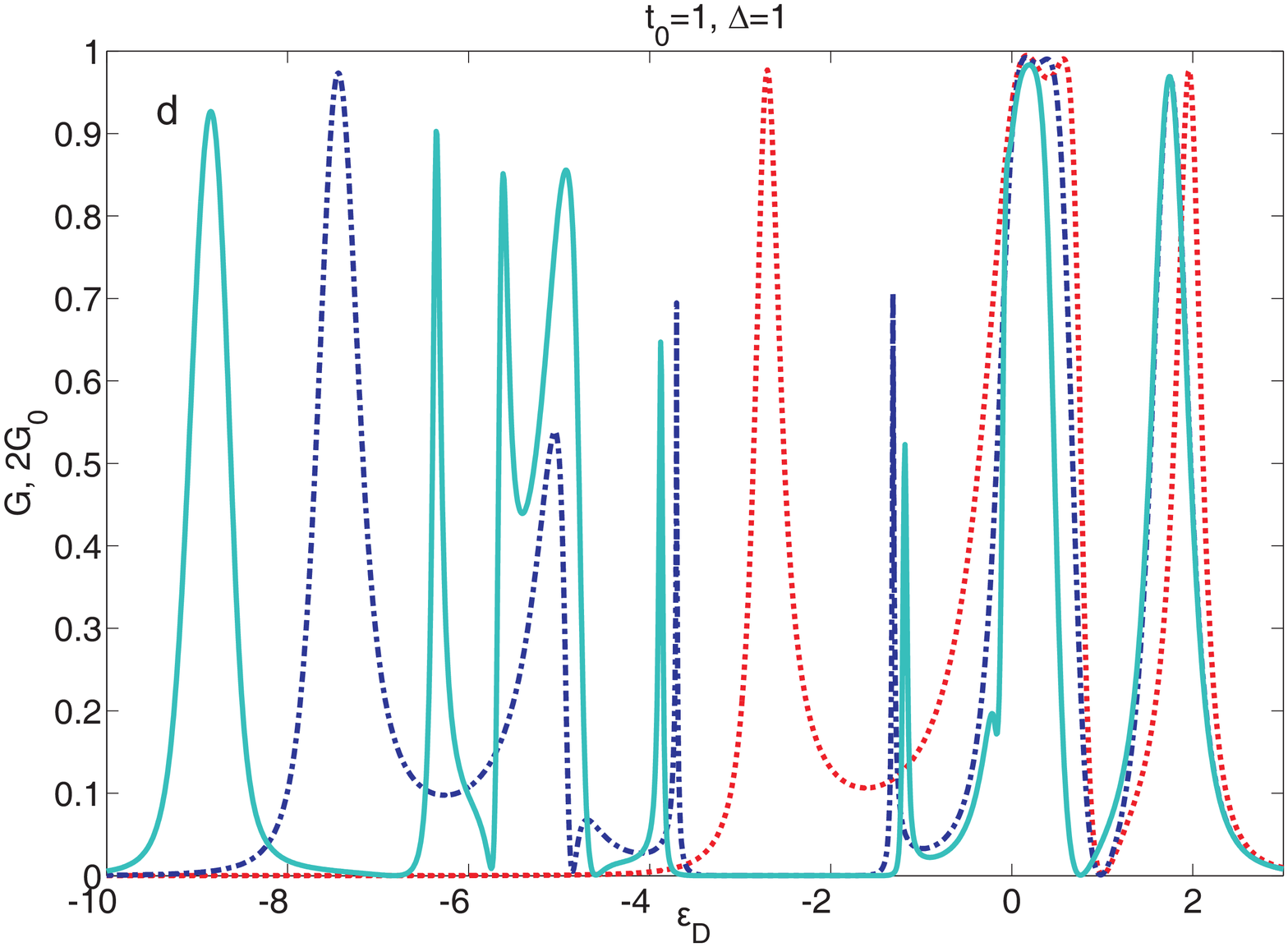}
\caption{The conductance of the isotropic QQD, $k_{B}T=0.01$: a) $t_0=\Delta=0$; b) $t_0=1,~\Delta=0$; c) $t_0=0,~\Delta=1$; d) $t_0=1,~\Delta=1$.} \label{6}
\end{center}
\end{figure}
The TPS splitting effect is not symmetrical. The modification of the left and right TPSs is different for $V\neq0$ (solid line at fig. \ref{6}a). The right TPS splitting leads to the broadening of the insulating band due to significant suppression of the first peak. If $U,~V\neq0$ and the carrier hopping between 2nd and 3rd QDs is activated the widths of the low conductance band and insulating band become even larger as it is depicted at figure \ref{6}b. As it has already discussed above the nonzero energy shift $\Delta$ induces the Fano antiresonance or the asymmetrical peak in the central part of the TPS (dotted line at fig. \ref{6}c). This antiresonance is doubled for $U\neq0,~V=0$ (dash-dot line) and the low conductance bands in the both TPSs appear without the interdot Coulomb correlations (the regions $\varepsilon_D=-8$ --- $-6.5$ and $\varepsilon_D=0.5$ --- $1.5$). Finally, $V\neq0$ results in two additional antiresonances (solid line) by analogy with the case $\Delta=0$. As a prominent result in the right TPS the antiresonance, $G\simeq0$, without the correlations (dotted line) is replaced by the resonance around $\varepsilon_{D}=0$, $G\simeq1$, with the correlations (solid line). If $t_{0},~\Delta=1$ one antiresonance moves from the right TPS to the left TPS (solid line at \ref{6}d).

\begin{figure}[h!] \begin{center}
\includegraphics[width=0.425\textwidth]{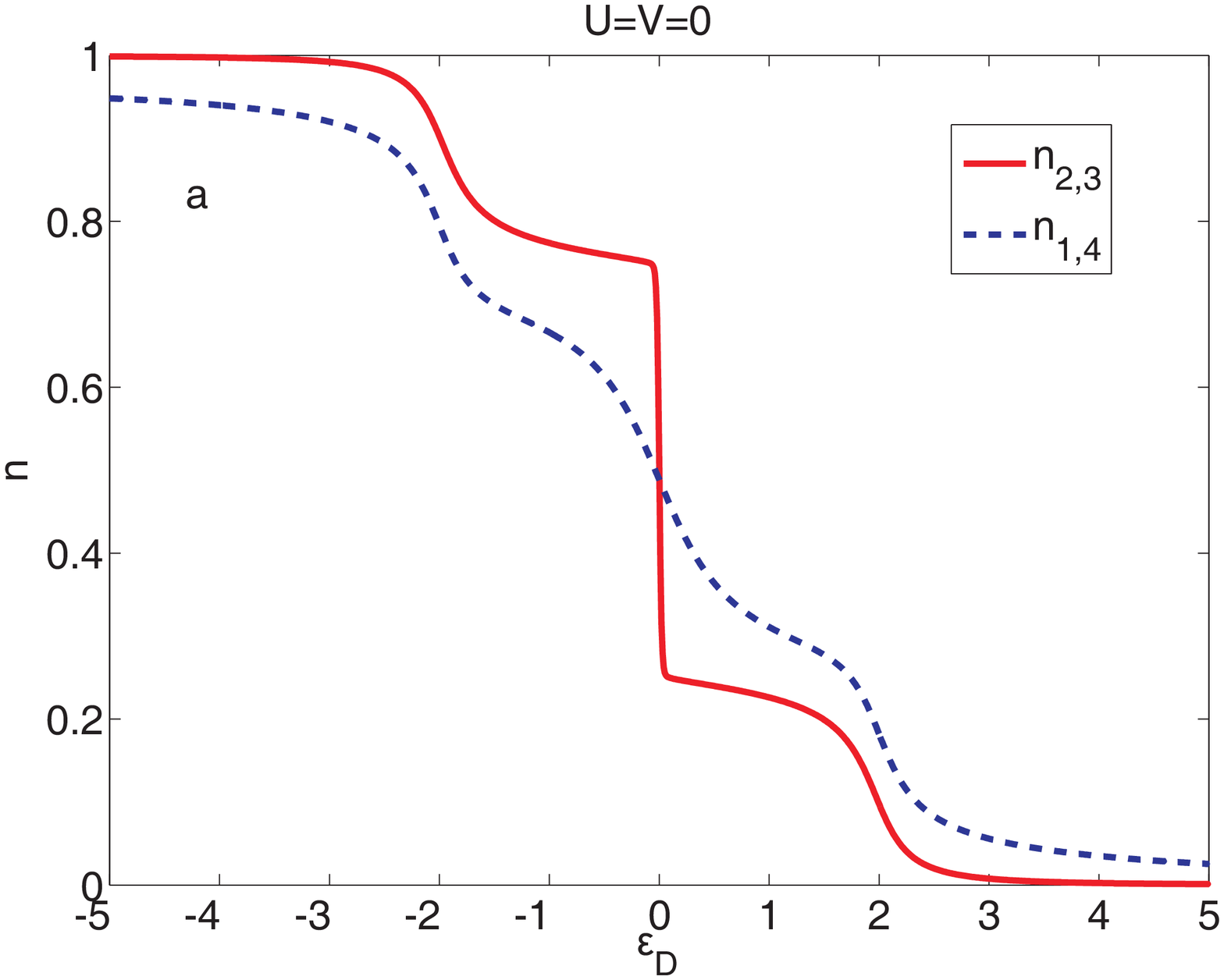}
\includegraphics[width=0.425\textwidth]{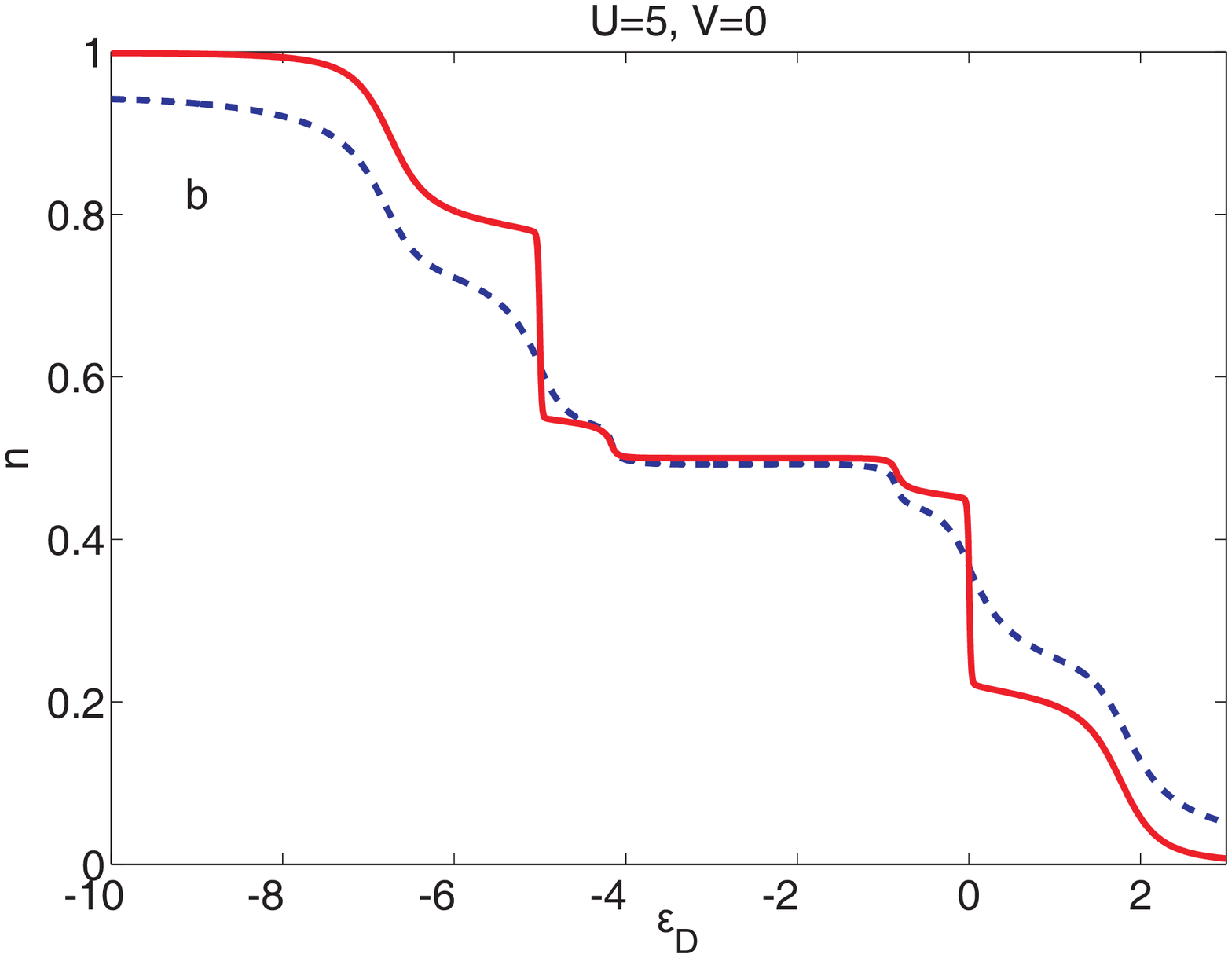}
\includegraphics[width=0.425\textwidth]{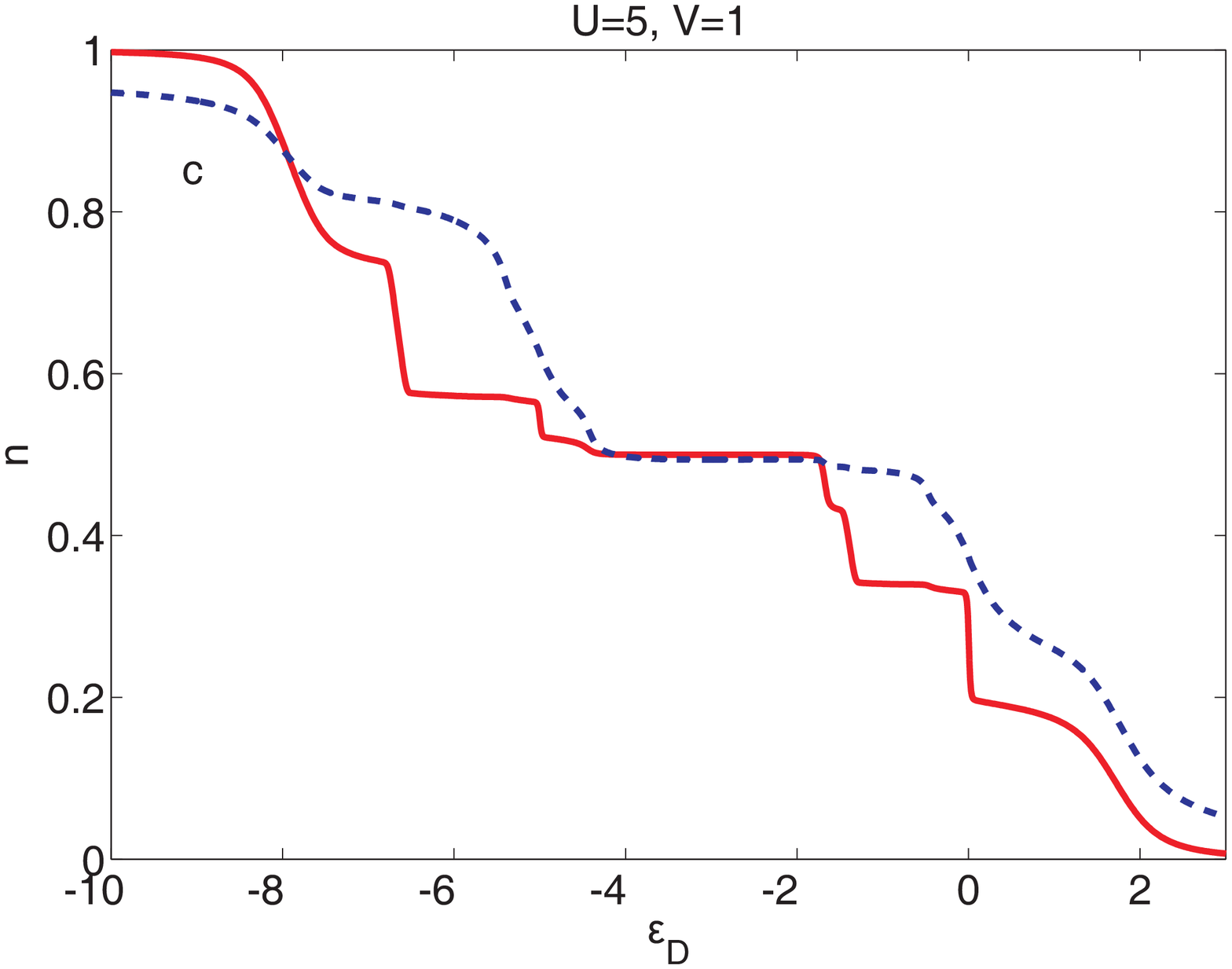}
\caption{The gate-field dependence of occupation numbers of the isotropic QQD, $t_{0}=\Delta=0$, $k_{B}T=0.01$: a) $U=V=0$; b) $U=5,~V=0$; c) $U=5,~V=1$.} \label{7}
\end{center}
\end{figure}
The described behavior of the conductance is determined by the corresponding features of the occupation numbers. It can be easily illustrated in the simplest regime, $t_{0}=\Delta=0$. In the absence of all the Coulomb interactions the gate voltage dependencies of the side, $n_{1,4}$, and the central, $n_{2,3}$, QD occupations have three steps at the same positions as the resonances in the TPS (see dashed and solid lines at fig.\ref{7}a). If the intradot correlations are taken into account this staircase obtains three more steps at a distance $U$ (fig.\ref{7}b). The interdot Coulomb interaction leads to the splitting of each central step of $n_{2,3}$ around $\varepsilon_{D}=-5,~0$ (fig.\ref{7}c) \cite{you-99a}. Importantly, the extensive areas where the occupations do not change correspond to the insulating bands and low conductance bands at figures \ref{5} and \ref{6}a. Lastly, the weak influence of the Coulomb correlations on the conductance at the high gate fields ($\varepsilon_{D} \geq 1$) is explained by the low occupation of the QQD's levels.

\begin{center}\textbf{B. Anisotropic QQD}\end{center}

\begin{figure}[h!] \begin{center}
\includegraphics[width=0.45\textwidth]{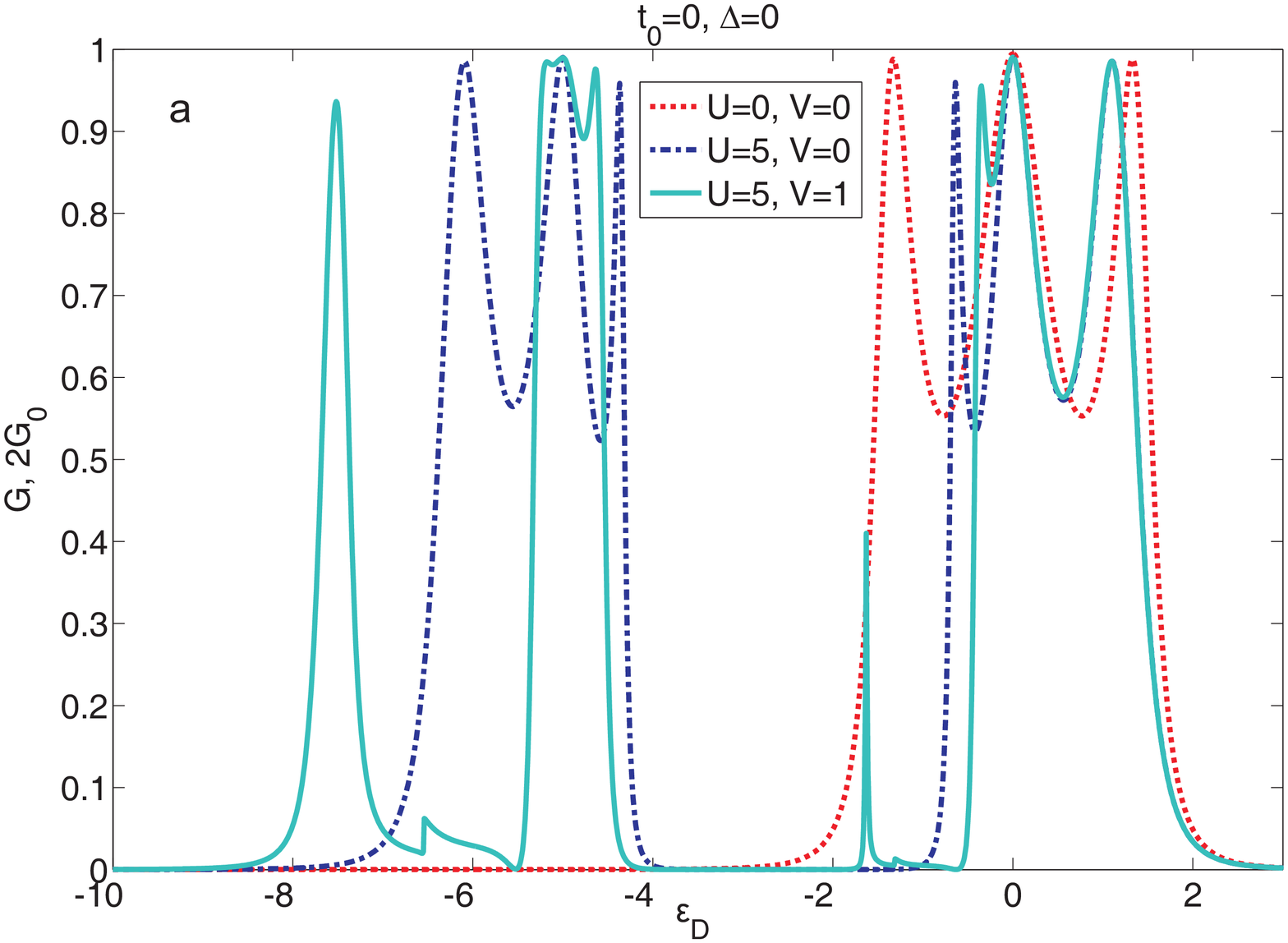}
\includegraphics[width=0.45\textwidth]{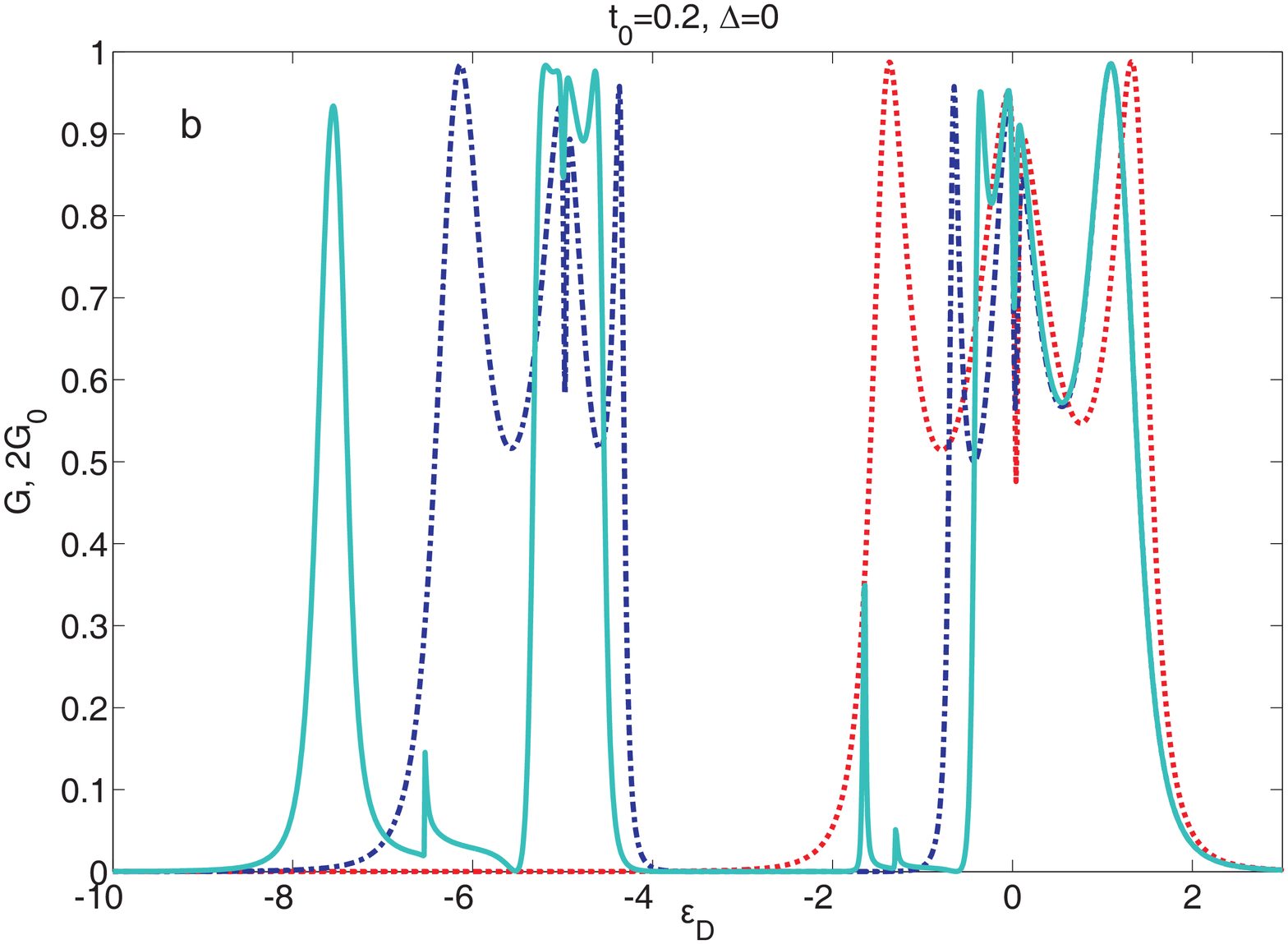}
\includegraphics[width=0.45\textwidth]{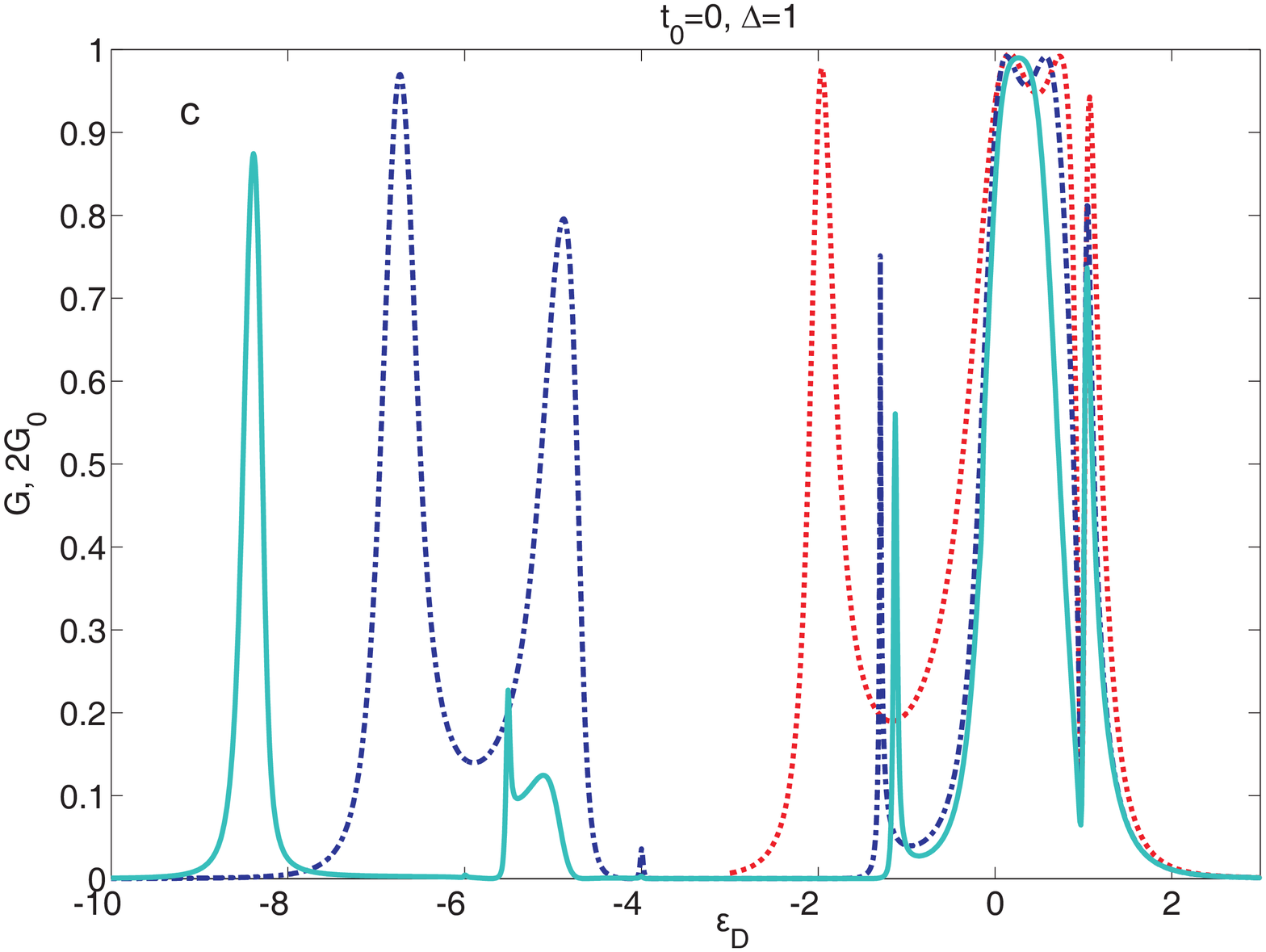}
\includegraphics[width=0.45\textwidth]{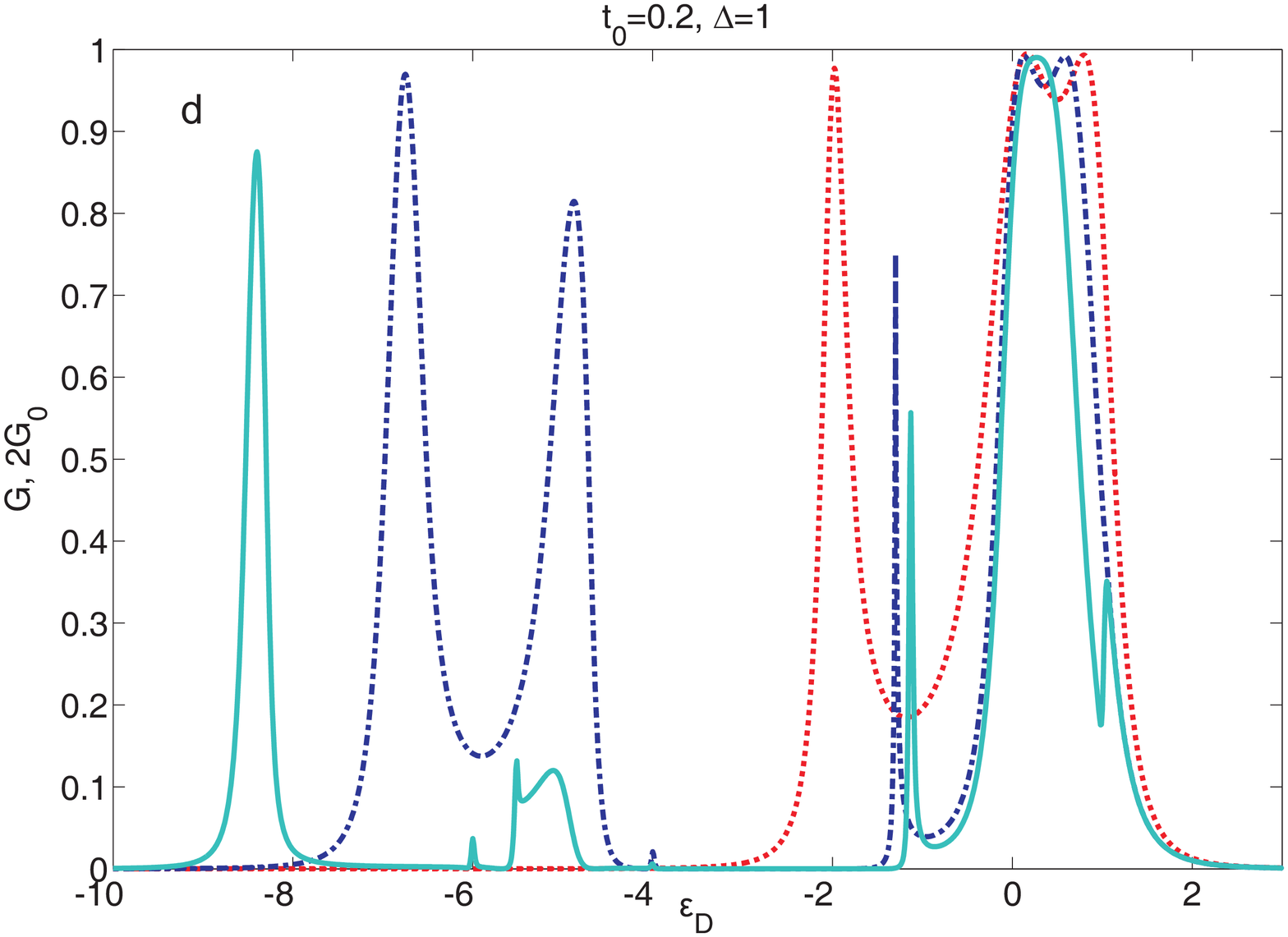}
\caption{The conductance of the anisotropic QQD, $t_{2}=0.1$, $k_{B}T=0.01$: a) $t_0=\Delta=0$; b) $t_0=0.2,~\Delta=0$; c) $t_0=0,~\Delta=1$; d) $t_0=0.2,~\Delta=1$.} \label{8}
\end{center}
\end{figure}

When the anisotropy of the hopping integrals takes place the effect of the interdot tunneling in the central part on the conductance is not strong in comparison with the isotropic situation. The influence of the Coulomb correlations results in the same features. However, the combination of the Coulomb interactions with the energy shift $\Delta$ can dramatically decrease the conductance in the left TPS (compare dash-dot and solid lines at figs.\ref{8}c,d). We can clearly see that the big insulating band occurs with small conductance peak emerging in the middle. Thus the EPE and gate fields can considerably modify the conductance by the significant suppression of the TPS.

\begin{center}\textbf{7. Conclusion}\end{center}

We have considered the spectral and transport properties of the QQD structure at low temperatures, low bias and the strong coupling regime. The treatment of the problem was based on the nonequilibrium Green's functions and the tight binding approximation. It is found that there are more than one way to observe the Fano effect in the system. First of them has already been mentioned before and consists of making two nonequivalent paths for electron waves by the energy shift $\Delta$ \cite{yan-13}. Additionally we showed that the anisotropy of the kinetic processes in the system, $t_{1} \neq t_{2}$, leads to the Fano-Feshbach asymmetrical peak for $t_{0}\neq0$ even though $\Delta=0$. The effect is explained in terms of resonant interaction between the bonding and antibonding states in the system. This scenario of the Fano effect seems to be more attractive for experimental observation since it does not need gate field. The anisotropy results in the dependence of the shape and width of the Fano-Feshbach resonance on the sign of $\Delta$ as well.

The problem of the influence of the Coulomb correlations on quantum transport in the QQD device was solved using the equation-of-motion technique for the retarded Green's functions. We applied the decoupling scheme of You and Zheng \cite{you-99a,you-99b} which allows to take into account the intra- and interdot Coulomb correlations beyond the Hartree-Fock approximation in nonmagnetic case. We demonstrated that the QQD structure has wide region of zero conductance with steep edges separating two TPSs if the intradot Coulomb interactions in each dot are allowed. This effect has been considered earlier for more sophisticated QD-based devices \cite{gong-06,fu-12}. The interdot Coulomb correlations between the central QDs results in the broadening of this band and the occurrence of the band with low conductance in the left TPS due to the Fano antiresonances. When the hopping between the central QDs is also permitted the bands become even wider. Furthermore, the conductance of the anisotropic QQD device can be remarkably modified by changing $\Delta$ if the interdot Coulomb repulsion is taken into account.

\begin{center}\textbf{Acknowledgment}\end{center}

We acknowledge fruitful discussions with P.I. Arseyev, N.S. Maslova, V.N. Mantsevich and R.Sh. Ikhsanov. This work was financially supported by the Comprehensive programme SB RAS no. 0358-2015-0007, the RFBR, projects nos. 15-02-03082, 15-42-04372, 16-42-243056, 16-42-242036. M.Yu. K. thanks the Program of Basic Research of the National Research University Higher School of Economics for support.

\end{document}